\documentclass[10pt,conference]{IEEEtran} 
\usepackage{cite}
\usepackage{amsmath,amssymb,amsfonts}
\usepackage{algorithmic}
\usepackage{graphicx}
\usepackage{textcomp}
\usepackage{xcolor}
\usepackage{textcmds}
\usepackage{enumerate}
\usepackage[hyphens,spaces,obeyspaces]{url}
\usepackage[multiple]{footmisc}
\PassOptionsToPackage{hyphens}{url}\usepackage{hyperref}
\usepackage{cleveref}
\usepackage[export]{adjustbox}
\usepackage{tcolorbox}
\usepackage{caption}
\usepackage{subcaption}
\usepackage{footnote}

\usepackage[font=small,labelfont=bf,tableposition=top]{caption}
\DeclareCaptionLabelFormat{andtable}{#1~#2  \&  \tablename~\thetable}

\def\BibTeX{{\rm B\kern-.05em{\sc i\kern-.025em b}\kern-.08em
    T\kern-.1667em\lower.7ex\hbox{E}\kern-.125emX}}
    
\usepackage{soul}
\usepackage{tikz}
\usetikzlibrary{calc}

\newcommand{\squeezeup}{\vspace{-2.5mm}}

\crefname{figure}{}{}
\Crefname{figure}{}{}

\makeatletter
\newif\if@anonymize

\@anonymizetrue    

\if@anonymize
  \newcommand{\highlight@DoHighlight}{
    \fill [outer sep = -15pt, inner sep = 0pt, color=black]
          ($(begin highlight)+(0,8pt)$) rectangle ($(end highlight)+(0,-3pt)$) ;
  }

  \newcommand{\highlight@BeginHighlight}{
    \coordinate (begin highlight) at (0,0) ;
  }

  \newcommand{\highlight@EndHighlight}{
    \coordinate (end highlight) at (0,0) ;
  }

  \newdimen\highlight@previous
  \newdimen\highlight@current
  \newlength{\item@width}

  \DeclareRobustCommand*\anonymize{%
    \SOUL@setup
    \def\SOUL@preamble{%
      \begin{tikzpicture}[overlay, remember picture]
        \highlight@BeginHighlight
        \highlight@EndHighlight
      \end{tikzpicture}%
    }%
    \def\SOUL@postamble{%
      \begin{tikzpicture}[overlay, remember picture]
        \highlight@EndHighlight
        \highlight@DoHighlight
      \end{tikzpicture}%
    }%
    \def\SOUL@everyhyphen{%
      \discretionary{%
        \SOUL@setkern\SOUL@hyphkern
        \SOUL@sethyphenchar
        \tikz[overlay, remember picture] \highlight@EndHighlight ;%
      }{%
      }{%
        \SOUL@setkern\SOUL@charkern
      }%
    }%
    \def\SOUL@everyexhyphen##1{%
      \SOUL@setkern\SOUL@hyphkern
      \settowidth{\item@width}{##1}%
      \makebox[\item@width]{}%
      \discretionary{%
        \tikz[overlay, remember picture] \highlight@EndHighlight ;%
      }{%
      }{%
        \SOUL@setkern\SOUL@charkern
      }%
    }%
    \def\SOUL@everysyllable{%
      \begin{tikzpicture}[overlay, remember picture]
        \path let \p0 = (begin highlight), \p1 = (0,0) in \pgfextra
          \global\highlight@previous=\y0
          \global\highlight@current =\y1
        \endpgfextra (0,0) ;
        \ifdim\highlight@current < \highlight@previous
          \highlight@DoHighlight
          \highlight@BeginHighlight
        \fi
      \end{tikzpicture}%
      \settowidth{\item@width}{\the\SOUL@syllable}%
      \makebox[\item@width]{}%
      \tikz[overlay, remember picture] \highlight@EndHighlight ;%
    }%
    \SOUL@
  }
\else
  \newcommand{\anonymize}[1]{#1}
\fi
\makeatother    

\begin{document}

\title{Revisiting Dockerfiles in Open Source Software Over Time}

\author{
\IEEEauthorblockN{Kalvin Eng}
\IEEEauthorblockA{\textit{Department of Computing Science} \\
\textit{University of Alberta}\\
Edmonton, Canada \\
kalvin.eng@ualberta.ca}
\and
\IEEEauthorblockN{Abram Hindle}
\IEEEauthorblockA{\textit{Department of Computing Science} \\
\textit{University of Alberta}\\
Edmonton, Canada \\
abram.hindle@ualberta.ca}
}

\maketitle

\begin{abstract}
Docker is becoming ubiquitous with containerization for developing and deploying applications. Previous studies have analyzed Dockerfiles that are used to create container images in order to better understand how to improve Docker tooling. These studies obtain Dockerfiles using either Docker Hub or Github. In this paper, we revisit the findings of previous studies using the largest set of Dockerfiles known to date with over 9.4 million unique Dockerfiles found in the World of Code infrastructure spanning from 2013-2020. We contribute a historical view of the Dockerfile format by analyzing the Docker engine changelogs and use the history to enhance our analysis of Dockerfiles. We also reconfirm previous findings of a downward trend in using OS images and an upward trend of using language images. As well, we reconfirm that Dockerfile smell counts are slightly decreasing meaning that Dockerfile authors are likely getting better at following best practices. Based on these findings, it indicates that previous analyses from prior works have been correct in many of their findings and their suggestions to build better tools for Docker image creation are further substantiated.
\end{abstract}






\section{Introduction}
Docker, a tool for creating and running programs in containers consistently across platforms, was initially released to the public on March 20, 2013~\cite{dockeryt,docker1}. Ever since its release, Docker has amassed a considerable following with 2.9 million desktop installations and 7 million Docker Hub users as reported in July 2020~\cite{dockeridx}.

The use of container software such as Docker has made applications easier to deploy, scale, and migrate across platforms. Furthermore, it has also made development setup simpler by reducing the amount of time needed to configure an appropriate environment by bundling the needed configuration instructions in a \textit{Dockerfile} which can then be used to create images for containers.

Because of the proliferation of Docker, this paper seeks to replicate and elaborate on previous studies on Dockerfile usage using the largest Dockerfile dataset~\cite{msr2021} known to date. This paper has findings, using data between 2013-2020, that include:
\begin{itemize}
    \item Discovering that 7.99\% of Dockerfiles exist in more than one distinct repository
    \item Most repositories overall contain up to 6 Dockerfiles
    \item Confirmation of previous findings such as JavaScript being the most popular language of projects that contain Dockerfiles~\cite{cito2017empirical,lin2020large} (2016, 2020) and {RUN} being the most popular Dockerfile instruction~\cite{cito2017empirical}
\end{itemize}

\section{Previous Work}
\label{prev-work}
In previous work, large collections of Dockerfiles have been mined from Github and Docker Hub to better understand Docker use in repositories and to gather insights on popularity, quality, and possible ways to improve Docker usage.

\subsubsection*{Mining Github} 
Cito et al.~\cite{cito2017empirical} (2016) focused on analyzing over 70,000 Dockerfiles in Github within commits up until October 2016 finding that: most Dockerfiles use heavy-weight operating systems as a base image; the biggest quality issue of Dockerfiles is missing version pinning of images; and Dockerfiles are not revised often.
In another study by Wu et al.~\cite{wu2020characterizing} (2020), 6334 projects were selected from Github and analyzed for Dockerfile smells finding that: 62\% of projects selected have code smells; newer and popular projects have less code smells; and projects with different languages have discernible differences in the amount of smells. 
Also of note is Henkel et al.~\cite{henkel2020learning} who retrieved approximately 178,000 Dockerfiles from Github to test with rules mined from the Dockerfiles of official Docker images and found that there should be more tooling to support developers using Dockerfiles.

\subsubsection*{Mining Docker Hub} 
Lin et al.~\cite{lin2020large} (2020) scraped Docker Hub and its related GitHub and Bitbucket repositories retrieving 434,304 Dockerfiles up until May 2020. They sought to better understand the Docker ecosystem through Docker Hub. They concluded that: for base images more programming runtime images and ready-to-use application images are being used instead of OS images; there is a declining trend over the years in Dockerfile smells; and there is an upward trend of using end of life Ubuntu base images. Additionally, Zhang et al.~\cite{zhang2018insight,zhang2019clustering} selected 2840 projects from Docker Hub to identify evolutionary patterns of Dockerfiles and its impact on Dockerfile quality and image build latency. It should be noted that mining from Docker Hub may not be representative of all Docker usage as users do not have to push images to Docker Hub to use Docker and can choose to build and host images locally or in a private repository.

\subsection{Challenges in Previous Work}
All of the above previous work focuses on Docker use in a project based perspective and involves mining Dockerfiles from git repositories on services like Github. In these services, projects are often cloned via \qq{fork} features or by manually copying other repositories leading to potential analysis of cloned work. Furthermore, files may be replicated among repositories that appear to be distinct.

In previous work, the heuristics used to choose projects containing Dockerfiles vary including counting the number of stars in Github for project popularity. We argue that by selectively choosing projects based on popularity, an unclear picture of how Dockerfile usage is created as projects can contain clones of Dockerfiles. Therefore, we do not use projects to select Dockerfiles but instead center our trend analysis around the unique Dockerfile blobs (identified by SHA-1s) committed year by year via git present in the \qq{World of Code}~\cite{ma2019world}.  

Furthermore, in previous studies analyzing Dockerfile code smells, the evolution of instruction use overtime in relation to the changes of the Docker engine itself over time were not considered. This paper introduces a comprehensive review of the annual evolution of the Dockerfile format in Section~\ref{dockerfile-format}.

\subsection{Replicating Previous Findings}
In light of the challenges in previous work, this paper aims to replicate the analysis of Dockerfiles in open source software on a year by year basis centered around unique Dockerfiles instead of projects with data up until September 2020 and intends to confirm or refute the following findings:
\begin{itemize}
    \item \qq{RUN} is by far the most popular instruction and often used to manage dependencies~\cite{cito2017empirical} (2016)
    \item Dockerfiles are not changed often~\cite{cito2017empirical} (2016)
    \item Most Dockerfiles use heavy-weight operating systems as a base image~\cite{cito2017empirical} (2016) which contrasts with most Dockerfiles using ready-to-use and application base images~\cite{lin2020large} (2020)
    \item The biggest Dockerfile quality issues are the lack of version pinning~\cite{cito2017empirical,lin2020large} (2016, 2020), newer and popular projects have less code smells~\cite{wu2020characterizing} (2020), and there is a declining trend over the years in Dockerfile smells~\cite{lin2020large} (2020)
\end{itemize}

Our intent to replicate previous studies mining large Dockerfile collections leads to the following research questions:
\begin{enumerate}[{RQ1.}]\bfseries
    \item How are instructions and base images used in Dockerfiles over time?
    \item How prevalent are code smells in Dockerfiles?
    \item How do Dockerfiles change over time?
\end{enumerate}

By revisiting previous studies of Dockerfile datasets with a significantly larger and newer dataset, we aim to provide a more comprehensive picture surrounding Dockerfile usage to better support high quality Docker image builds and better tooling within the Docker ecosystem.







\section{Dockerfile Data}
\label{sec:data-retrieval}
The Dockerfile data is obtained from the S version of WoC (World of Code)~\cite{ma2019world}. The latest S version contains 9,192,143,411 unique blobs, 2,326,066,436 commits, and 135,162,320 distinct repositories collected from open source communities including  GitHub, Bitbucket, and GitLab identified on August 28, 2020 and retrieved by September 18, 2020~\cite{wocstats2020}. A \textit{distinct repository} is determined as the \qq{most central} repository to represent a group of repositories found with the Louvain community detection algorithm~\cite{Mockus_2020_P}. By using distinct repositories, many cloned projects (forks) can be avoided when performing an analysis.

\subsection{Obtaining Data}
Dockerfiles are found by identifying the SHA-1s of blobs with the filename containing \qq{Dockerfile}  (non case-sensitive). We do not check for an exact match for a filename containing only \qq{Dockerfile} as other phrases can be prepended and appended to \qq{Dockerfile} and still be used to build a container image~\cite{so_docker_naming}. From this initial search, 12,201,624 filenames mapping to blob SHA-1s are obtained.

A unique list of blob SHA-1s with 10,167,327 related to filenames containing \qq{Dockerfile} (non case-sensitive) are obtained from the initial 12,201,624 blob SHA-1s. It should be noted that 15 blob SHA-1s are omitted as they are linked to over 100,000 or more commits in repositories and clearly do not contain any Dockerfile-like data.\footnote{\href{https://github.com/ssc-oscar/lookup/blob/5e78bbfe322a83f425c2cf8d7982d2be4e82d79b/woc.pm\#L718-L812}{https://github.com/ssc-oscar/lookup/blob/5e78bbf/woc.pm/\#L718-L812}}

We obtain the contents of 10,065,114 blobs from the 10,167,327 blob SHA-1s. Some of the blob contents are missing since WoC only stores the contents of non-binary files as classified by the libgit2 library \textit{git\_blob\_is\_binary} function~\cite{libgit2}.

We parse the blob contents using the Python dockerfile library~\cite{pydockerfile} (a wrapper for the official Dockerfile parser) and store the returned metadata for each command. It should be noted that the contents of the blobs are encoded to be \mbox{UTF-8} with null chracters removed and any unknown characters replaced with a backslashed escape sequence as handled by Python 3~\cite{unicode}. 

During the parsing of blob contents, another 435,490 blobs are removed due to three cases: not being parseable, not containing \qq{FROM} as the first instruction, or not having \qq{ARG} precede \qq{FROM} as the initial instructions. For the first case there are 17,718 blobs and for the latter cases there are 417,772 blobs.
With these three cases omitted, a unique list of 9,629,624 parseable blob SHA-1s and contents is retrieved. The parseable blobs are broken down into 117,819,113 lines of Dockerfile instructions via the Docker parser with 1,202,952 being invalid instructions. In total, there are 9,445,029 blobs with all valid instructions.

Of the 9,629,624 blob SHA-1s, 11,515,615 unique commits are found to be linked to the blob SHA-1s. These unique commits are linked to 1,950,994 \textit{distinct repositories} that represent a group of repositories. Overall, the unique commits exist in 4,470,158 repositories. It should be noted that 11,706 commits could not be linked to repositories as they may have become orphaned by the time of mining and were unable to be linked to a repository. Of the 4,470,158 repositories we find that commits are present in 4,279,731 repositories on Github, 121,187 on GitLab, 65,247 on Bitbucket, 1620 on Debian Salsa Gitlab, 2079 on GNOME GitLab, and 294 on other open source git version control systems. 

For the distinct repositories, additional available metadata information including the types of files, the number of files, the community size determined with the Louvain method, the number of commits, and the initial commit date is retrieved from the WoC MongoDB for the S version for 1,950,990 repositories as of December 2020. Four repositories did not have metadata since their commits did not contain valid author timestamps.

All of our data mined from the WoC is inserted into a Postgresql database 
for querying to provide an overview of the state of Dockerfiles in open source software in Section~\ref{df_state} and to also answer the research questions in Section~\ref{RQ}. We also provide a replication package to reproduce our work~\cite{msr2021}. To our knowledge, this is the most extensive set of Dockerfiles created to date.


\subsection{Use of Data for Analysis} 
Because the data is mined from such a vast array of sources, the data may be inaccurate as some author timestamps are unreasonable (e.g.\ having a commit on January 1, 1970). Therefore, we are particularly interested in blobs committed from the start of 2013 until 2020 to answer our research questions. Within the time frame there are 9,456,011 unique blob SHA-1s that have been committed and are parseable. These SHA-1s are linked with 1,950,964 distinct repositories overall.
We filter the blobs further as needed since the author timestamps of the commit containing the Dockerfile may be inaccurate (blob is created on a different date). To avoid confusion as to which blobs are used for analysis, we repeat the counts of SHA-1s and distinct repositories throughout the paper.



\section{The State of Dockerfiles}
\label{df_state}
Since the WoC allows for the linking of blobs to commits and commits to projects, it allows for an analysis to determine the number of unique Dockerfile blobs present in each distinct repository. It also allows for an analysis to see how Dockerfiles may be cloned among distinct repositories. WoC also classifies the languages of distinct repositories allowing us to revisit previous studies of the distribution of programming languages in projects. This section presents some novel insight into how Dockerfiles relate with communities of repositories in open source software.

\subsection{Dockerfile Commit Ratio}
\label{df_ratio}
Upon counting the blobs and commits for each distinct repository, it becomes evident that some repositories contain a significant amount of Dockerfile blobs. 
Thus, we introduce a simple heuristic to determine how many blobs on average a repository may contain by taking the total count of Dockerfile blobs related to commits of a distinct repository and dividing it by the total count of Dockerfile commits of a distinct repository. This heuristic can be used to help identify anomalies where a repository is housing an atypical number of Dockerfiles as the ratio approximately corresponds to the number of Dockerfiles in a repository.

Using Dockerfile commits between the start of 2013 until 2020 in 1,950,964 distinct repositories, we find that 99\% of distinct repositories contain a rounded ratio of 0-6 which can be seen broken down in Table~\ref{tab:ratio-percent}. In terms of descriptive statistics for the ratio, we get a standard deviation of 66.94, a mean of 1.39, and a maximum of 89,110. A ratio can be 0 since a blob with the same SHA-1 might be committed in multiple commits of the same distinct repository creating a ratio close to 0.   

\begin{table}[hbtp]
\centering
\caption{Rounded ratio proportions of distinct repositories.}
\label{tab:ratio-percent}
\tiny
\begin{tabular}{|l|c|c|c|c|c|c|c|c|}
\hline
\textbf{Rounded Ratio}               & 1    & 2    & 0    & 3    & 4    & 5    & 6    & Other \\ \hline
\textbf{\% of Distinct Repositories} & 78.6 & 9.24 & 6.61 & 2.47 & 1.22 & 0.51 & 0.34 & 1.01  \\ \hline
\end{tabular}
\squeezeup
\end{table}

Of the 1\% (Other) of the rounded ratios, we present the top 6 projects in Table~\ref{tab:ratio-top} and bottom 6 projects in Table~\ref{tab:ratio-bottom}. It should be noted that some distinct repositories no longer exist, therefore an attempt was made to find a live online version related to the distinct repository community using the \qq{P2p} mapping in WoC; the links to the related existing repository are present as citations in the tables.

With manual inspection of the top 6 distinct repositories in Table~\ref{tab:ratio-top} that were able to be accessed, it is found that these repositories host many Dockerfiles as libraries for research, archiving, and to deploy software.

\begin{savenotes}
\begin{table}[tbp]
\centering
\caption{Top 6 distinct repositories of Other in Table~\ref{tab:ratio-percent}.}
\label{tab:ratio-top}
\resizebox{\linewidth}{!}{%
\begin{tabular}{|l|l|l|l|l|}
\hline
\textbf{Distinct Repository} & \textbf{Blobs} & \textbf{Commits} & \textbf{Ratio} & \textbf{Purpose} \\ \hline
irvin-s\_docker\_repair~\cite{irvin-s}      & 178221         & 2                & 89110          &  \vtop{\hbox{\strut Research using the extracted dataset of}\hbox{\strut Henkel et al.~\cite{henkel2020learning}}}                 \\ \hline
x0rzkov\_dockerfiles-search~\cite{x0rzkov}  & 232608         & 12               & 19384          & Mined Docker Hub images                 \\ \hline
volt72\_dockerfiles ~\cite{volt72}         & 55609          & 4                & 13902          & 
\vtop{\hbox{\strut Generate Dockerfile metadata}\hbox{\strut from Docker Hub images}}      \\ \hline
vsoch\_dockerfiles~\cite{dino-docker}           & 213154         & 18               & 11842          & Mined Docker Hub images~\cite{sochat_dino}                 \\ \hline
lonroth\_woocommerce-docker~\cite{woocommerce-docker}  & 20004          & 3                & 6668           & Deploy WooCommerce WordPress                 \\ \hline
lizebang\_docker-images      & 26846          & 5                & 5369           &     Could not be retrieved             \\ \hline
\end{tabular}%
}
\end{table}
\end{savenotes}

In comparison, with the manual inspection of the bottom 6 distinct repositories in Table~\ref{tab:ratio-bottom}, it is found that most of these repositories use Dockerfiles to run software in a new environment. 

\begin{table}[tbp]
\centering
\caption{Bottom 6 distinct repositories of Other in Table~\ref{tab:ratio-percent}.}
\label{tab:ratio-bottom}
\resizebox{\linewidth}{!}{%
\begin{tabular}{|l|l|l|l|l|}
\hline
\textbf{Distinct Repository}     & \textbf{Blobs} & \textbf{Commits} & \textbf{Ratio} & \textbf{Purpose} \\ \hline
laarid\_docker~\cite{laarid}                   & 501            & 77               & 7              &  Run android apps in Debian                \\ \hline
sachanda\_docker-ansible-1~\cite{dock-ansible}       & 860            & 132              & 7              &  Run ansible in different OSes                \\ \hline
Mpit4365\_teamcity~\cite{teamcity}               & 261            & 40               & 7              & Run TeamCity CI/CD Server                 \\ \hline
fouadsemaan\_docker-ansible-role~\cite{dock-ansible2} & 98             & 15               & 7              & Run ansible in different OSes                 \\ \hline
chrsm\_harbor~\cite{theharbor}                    & 72             & 11               & 7              & Personal Docker image library                 \\ \hline
silky\_i2kit~\cite{i2kit}                     & 118            & 18               & 7              &   Deploy Linux containers               \\ \hline
tomeliason\_d98821~\cite{eliason}               & 59             & 9                & 7              &  Practice coursework                \\ \hline
defn\_aws-service-operator~\cite{aws}       & 105            & 16               & 7              &    Manage Amazon Web Services              \\ \hline
\end{tabular}%
}

\end{table}

We also determine that 6815 distinct repositories with a ratio greater than 6 have 1,474,366 distinct blobs not present in other repositories, making up 15.3\% of 9,620,453 Dockerfiles present in the dataset. Of particular interest is \qq{arpl\_base-images}~\cite{balena} (which is linked to an IoT Docker image library) containing 936,440 (63.51\%) of the 1,474,366 distinct blobs not present in other repositories. In addition, we find that the top 6 distinct repositories of Table~\ref{tab:ratio-top} contain another 201,415 (13.66\%) of the 1,474,366 distinct blobs not present in other repositories.

The results of the distinct repositories found using the ratio suggests that a large proportion of Dockerfiles in our mined data exists in a small number of projects. However, we should not remove the Dockerfiles present in these distinct repositories from the dataset since many Dockerfiles are mined from Docker Hub signifying that they have been useful to at least the person who pushed the image to Docker Hub. Furthermore, real world deployments may use Dockerfiles in the repository like the IoT Docker image library~\cite{balena}. In the next subsection, we examine how Dockerfiles can exist among many distinct repositories.

\subsection{Most Cloned Dockerfiles}
\label{df_cloned}
We obtain the distinct repository count and commit count for 9,456,011 unique blob SHA-1s. For the distinct repository count we find that most blobs exist in 1 distinct repository with a mean of 1.24 repositories and a standard deviation of 16. In comparison, we find that the commit counts have a mean of 2.6 commits with up to 75\% of blobs having 2 commits and up to 50\% of blobs having only 1 commit. It should be noted that there is a large variance for commit counts with a standard deviation of 1179. We compare the distributions of counts using the Kruskal-Wallis test and find p$<$0.001 and therefore commit counts are not necessarily linked to distinct repository counts.



We find that 768,508 Dockerfile blobs (7.99\%) exist in more than one distinct repository. As we wish to see what kinds of Dockerfiles are the most prevalent throughout open source software, we observe the top 10 cloned blobs present among distinct repositories in Table~\ref{tab:cloned-df}. In the table, we note the blob SHA-1, distinct repository count, commit count, year ($\geq$ 2013) first committed, base image, and the blobs interpreted purpose based on manual inspection of the Dockerfile. It should be noted that for some blobs there were multiple commit dates, hence we take the oldest date greater than or equal to 2013 (the year when Docker was first introduced) as the year.

\begin{table}[tbp]
\centering
\caption{Top 10 cloned Dockerfiles.}
\label{tab:cloned-df}
\resizebox{\linewidth}{!}{%
\begin{tabular}{|l|r|r|r|l|l|}
\hline
\textbf{Blob SHA-1}                           & \textbf{Repositories} & \textbf{Commits} & \textbf{Year} & \textbf{Base Image(s)}     & \textbf{Purpose}   \\ \hline
\tiny{990c12e0f20e21ba917587868f6a452c0d4f9d64} & 24859                 & 28991            & 2015          & ubuntu:12.04               & Run JavaScript     \\ \hline
\tiny{264c5c036d03f3b9e53f0e2405fdca00b37d9d24} & 20960                 & 23890            & 2018          & php:5.6-cli                & Run PHP            \\ \hline
\tiny{53fc52579744539dc9d223be13fe7be565396bf2} & 16830                 & 29280            & 2014          & ubuntu:trusty              & Build and Run curl \\ \hline
\tiny{f6a095230e85638cc9f98dd7beef2bcf8c87e98e} & 8512                  & 10357            & 2019          & composer:latest, php:7.3   & Run PHP Guzzle     \\ \hline
\tiny{c14d9894645da81ce14370f77002850fbee6a504} & 8337                  & 9622             & 2018          & ubuntu:latest              & Build node-sqlite3 \\ \hline
\tiny{5a19c7311639bc2186aa2501f921cc01df4aab93} & 8308                  & 9585             & 2019          & arm64v8/node:carbon        & Build node-sqlite3 \\ \hline
\tiny{4543daff8c4bb3b132a326ee892bf9d043d4a56b} & 8117                  & 10236            & 2017          & node:6                     & Run sequelize      \\ \hline
\tiny{ed587a774f09569b1f4487906a0599ffc8586823} & 7580                  & 8692             & 2019          & resin/rpi-raspbian:stretch & Build node-sqlite3 \\ \hline
\tiny{5b1552375393d83596edfa6d5edaebdf3f2d4261} & 7099                  & 7188             & 2018          & openjdk:8-jdk-slim         & Run Java jar       \\ \hline
\tiny{42e341df485d0d96252c69f306f894117ea6b953} & 6424                  & 7371             & 2017          & ubuntu:14.04               & Example Dockerfile \\ \hline
\end{tabular}%
}
\squeezeup
\squeezeup
\end{table}

With the manual inspection of the top 10 most cloned Dockerfiles, we find that the main purpose of Dockerfiles is to build popular software or be a base image to run software in a programming language.


\subsection{Programming Language Distribution}
\label{pl_dist}
In previous related work, the top 15 primary languages of Dockerfile containing projects in 2016~\cite{cito2017empirical} and top 12 primary languages of Dockerfile containing projects from 2013-2020~\cite{lin2020large} are found to have similar popular programming languages. We seek to confirm this finding using the primary programming languages of distinct projects determined by the WoC infrastructure.

\begin{figure}[tbp]
\centering
\includegraphics[width=\linewidth]{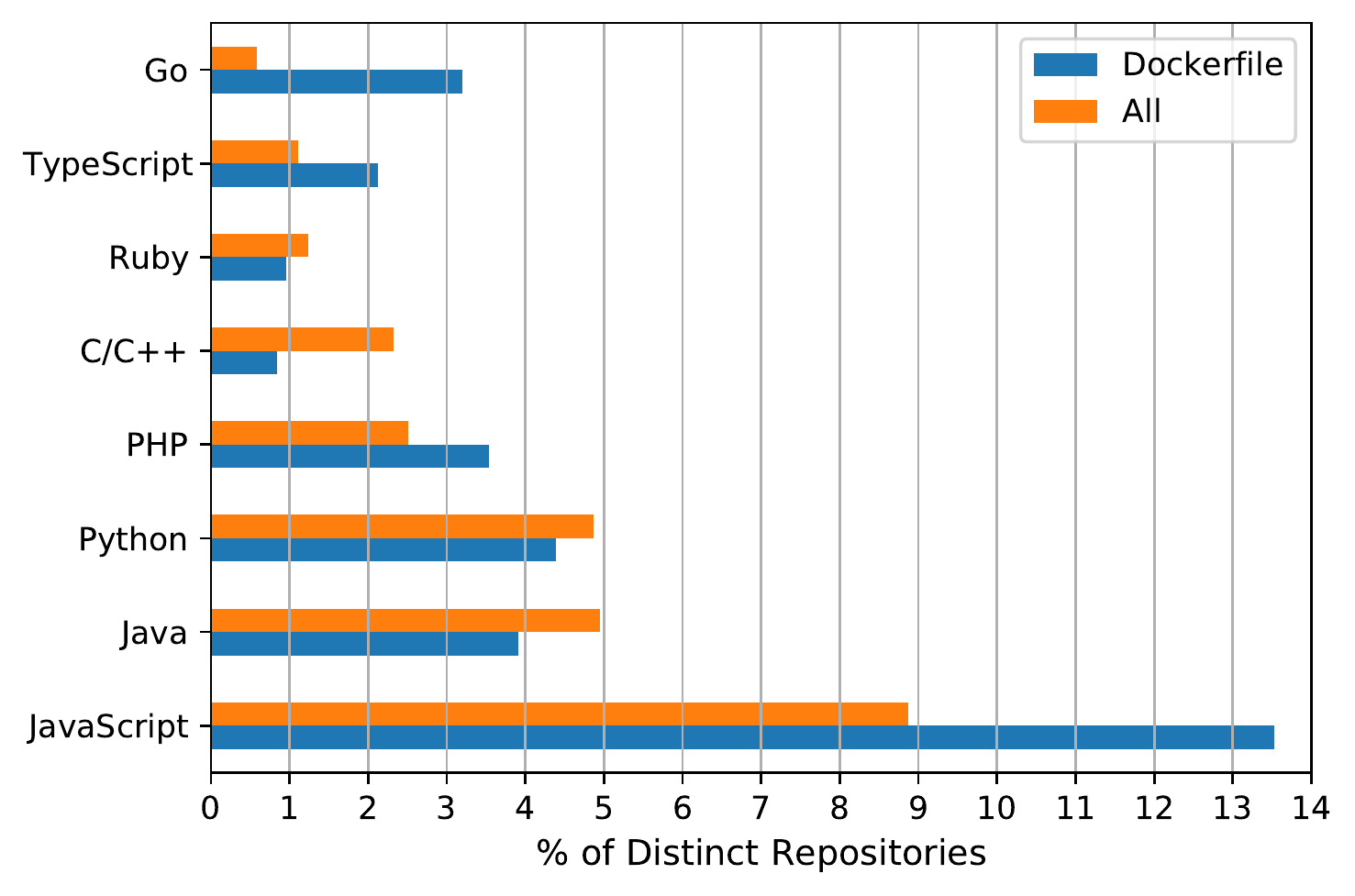}
\caption{Top 8 primary languages of Dockerfile containing distinct repositories and all WoC distinct repositories in 2020.}
\label{fig-lang-pop}
\squeezeup
\squeezeup
\end{figure}

In Figure~\ref{fig-lang-pop}, we present the top 8 primary programming languages in Dockerfile repositories and all of WoC. We find that JavaScript is more popular among Dockerfile repositories compared to all the repositories of WoC. We also see a similar pattern for Go, PHP, and Typescript. Notably, there are substantially less Dockerfile repositories with C/C++ compared to all of WoC. This suggests that repositories containing Dockerfiles tend to use more mainstream programming languages like JavaScript, PHP, Typescript, and Go.

With the WoC data including 1,950,990 distinct repositories containing Dockerfiles, we find that the top 8 languages presented in Figure~\ref{fig-lang-pop} to be present in the top languages found by Cito et. al~\cite{cito2017empirical} and Lin et al.~\cite{lin2020large}. However, we note that WoC only classifies 28 distinct languages based on file extensions, while GitHub classifies projects using Linguist~\cite{linguist} with the ability to classify 573 (at the time of writing) kinds of programming languages. As well, WoC does not classify shell scripts, while previous studies did~\cite{cito2017empirical, lin2020large}. Therefore, this should not be considered as a one-to-one comparison.

Interestingly, it appears that JavaScript and Python are both quite popular in our findings and previous studies. However, we do find that Go with WoC classification is less popular compared to previous studies which may be due to how WoC classifies the programming languages of a repository.

 \noindent \hrulefill
 
 \noindent \textit{\textbf{State of Dockerfiles:}} Most repositories that contain Dockerfiles, contain up to 6 Dockerfiles and only a small portion of Dockerfiles are cloned among different distinct repositories. For projects that contain Dockerfiles, they are most likely to be coded in JavaScript.
 
\noindent \hrulefill





\section{Dockerfile Format Evolution}
\label{dockerfile-format} 
Over the years, the Docker tool has added and deprecated instructions in the Dockerfile specification. 
Due to the evolving nature of Dockerfile specifications, caution should be taken when analyzing how the Docker ecosystem has evolved. For example, a code smell defined by the Dockerfile linter finds that using \qq{MAINTAINER} is a smell since it was deprecated in 2017. However, all Dockerfiles created before 2017 should not consider using \qq{MAINTAINER} as a code smell for analysis as it was considered acceptable at the time. As such, we introduce the changes made in the Dockerfile specification over time by reviewing the Docker changelog~\cite{docker_rn} to aid our analysis in Section~\ref{RQ}.


\subsubsection*{2013}
The concept of a Dockerfile to create docker images was first introduced into the Docker tool on April 11, 2013 about one month after revealing Docker 0.1.4 to the world~\cite{hykes_dockerbuild}. In the initial concept, only the instructions \qq{FROM}, \qq{RUN}, and \qq{COPY} were used as seen in the initial example~\cite{hykes_changefile}. In the subsequent weeks, the concept was christened to be a \textit{Dockerfile}~\cite{hykes_hackdocker}. Dockerfile was first officially documented in version 0.3.1 to have the \qq{FROM}, \qq{MAINTAINER}, \qq{RUN}, \qq{CMD}, \qq{EXPOSE}, \qq{ENV}, and \qq{INSERT} commands. The \qq{ADD} command was also subsequently added in version 0.3.4. It should be noted that \qq{INSERT} was quickly deprecated in favor of \qq{ADD} in version 0.4.5. Other instructions were also introduced including: \qq{ENTRYPOINT} in version 0.4.8, \qq{VOLUME} in version 0.5.0, and \qq{USER} and \qq{WORKDIR} in version 0.6.0.


\subsubsection*{2014} 
The \qq{ONBUILD} instruction was added in version 0.8.0 so that instruction execution in a base image could be deferred until the time of building a subsequent Docker Image. In addition, the \qq{COPY} instruction was also introduced in version 0.12.0 so that files could be copied without extraction.

\subsubsection*{2015} 
The \qq{STOPSIGNAL} instruction was added in version 1.9.0 so that different syscalls could be specified for terminating processes in an image. The \qq{ARG} instruction was also introduced in version 1.9.0 so that variables could be passed to the builder from the command line at build time. As well, starting from version 1.5.0, \qq{FROM scratch} was made to be interpreted as a no-base specifier, and files not named \emph{Dockerfile} could be built by specifying a -f flag in the command line.

\subsubsection*{2016} 
The \qq{LABEL} instruction was introduced in version 1.11.1 to add metadata to the image being built. In version 1.12.0, two more instructions were also introduced: the \qq{HEALTHCHECK} instruction allowing for a command with an exit code to run to detect a container status and the \qq{SHELL} instruction to change the default shell in \qq{RUN} instructions. It should also be noted that the \qq{\#escape=} directive was also added to change the default \qq{\textbackslash} escape character so that Windows paths could be defined in Dockerfiles.

\subsubsection*{2017 and Onward} In 2017, the \qq{MAINTAINER} instruction was deprecated in version 1.13.0. In version 18.03.0-ce, Dockerfiles no longer needed to be within a build-context to build an image. Also of note was introducing the support for multi-stage builds in version 17.05.0-ce. In 2018, version 18.06.0-ce introduced a new optional experimental backend based on BuildKit allowing for images from external \mbox{Dockerfile} implementations to be built. 



\section{Trends Over Time}
\label{RQ}
We perform an empirical analysis in terms of trends annually from 2013-2020 to better understand how the Docker ecosystem is evolving over time. In this section we drill down and analyze how Dockerfiles change over time, what kinds of instructions and base images Dockerfiles use, and also any possible Dockerfile best practices that may or may not be followed.

To determine if there is any significance among 
distributions over the years for the types of bases images, number of smells, and number of revisions, we perform the Kruskal-Wallis (KW) test. If significant with two pairs or more, we perform the Mann-Whitney (MW) rank test with Holm adjustment to determine the significance among the distributions. The null hypothesis for KW test and MW rank test is that the distributions are not significantly different among each other. We reject the null hypothesis if p$<$0.05.

\subsection*{{\normalfont\textbf{RQ1. How are instructions and base images used in Dockerfiles over time?}}}
\begin{figure}[tbp] 
    \centering
  \subfloat[Decreasing standard instructions\label{subfig:dec}]{%
       \includegraphics[width=0.5\linewidth]{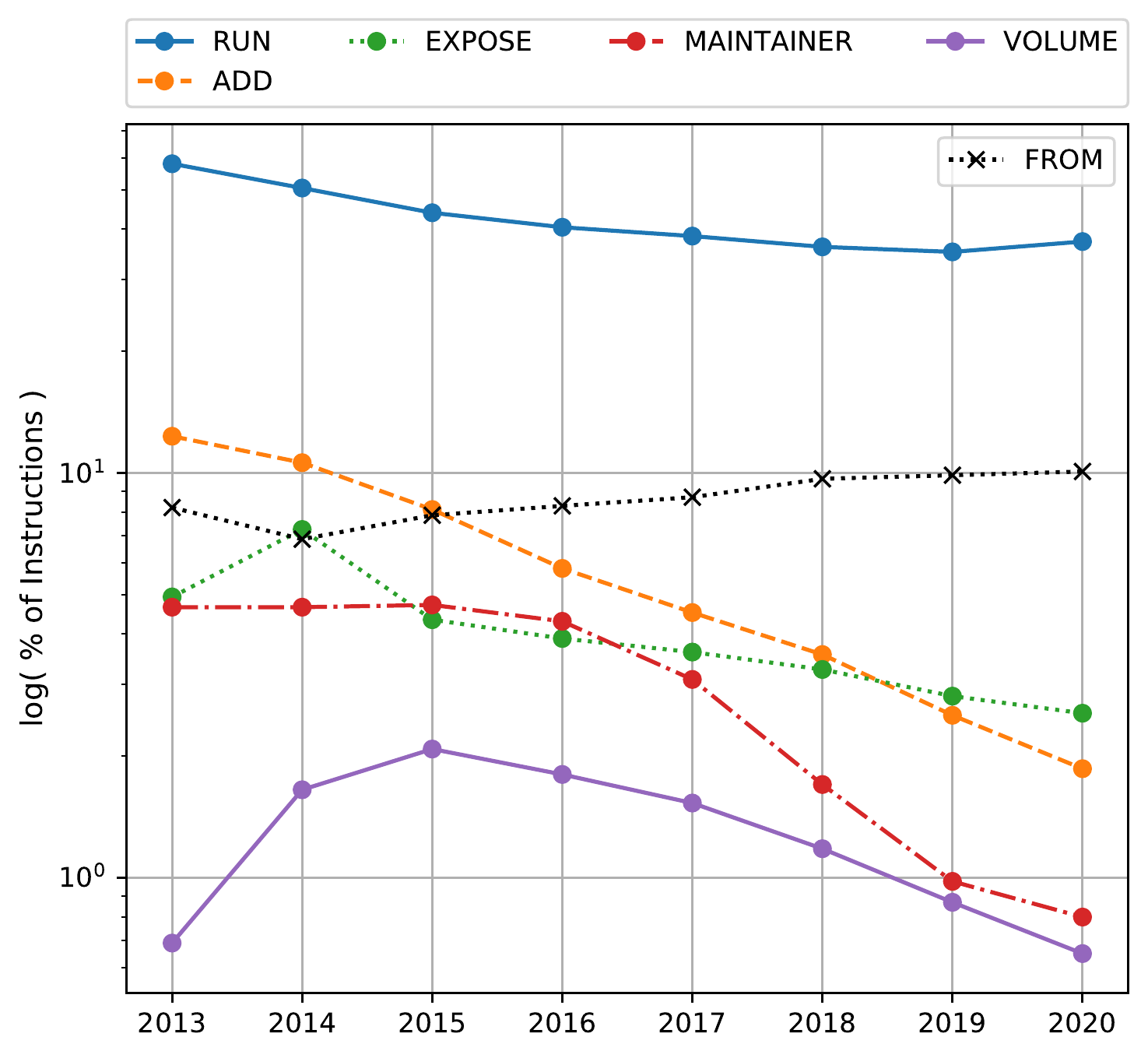}}
    \hfill
  \subfloat[Increasing standard instructions\label{subfig:inc}]{%
        \includegraphics[width=0.5\linewidth]{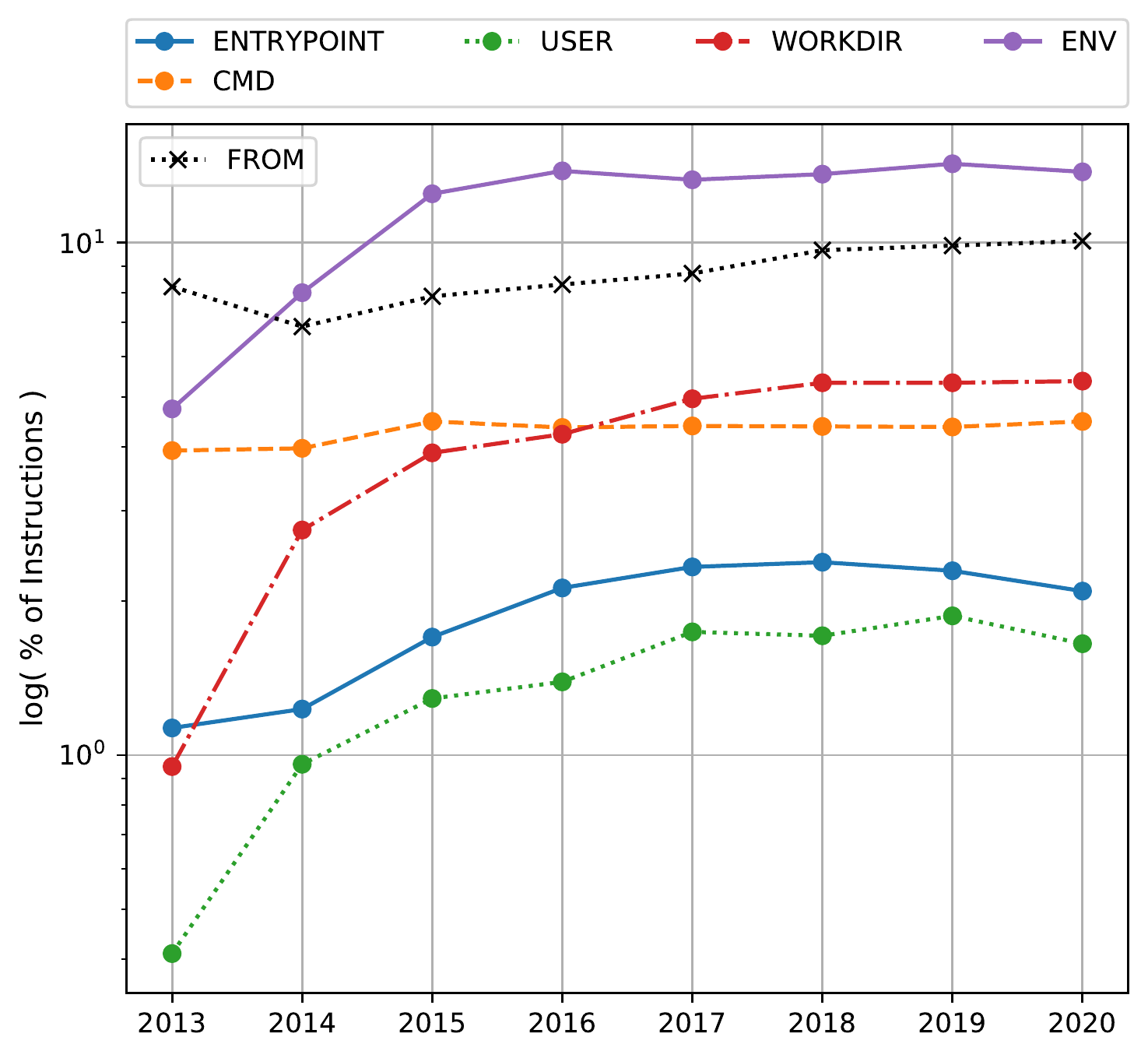}}
        \\
  \subfloat[New instructions up to 12\%\label{subfig:gt10}]{%
        \includegraphics[width=0.5\linewidth]{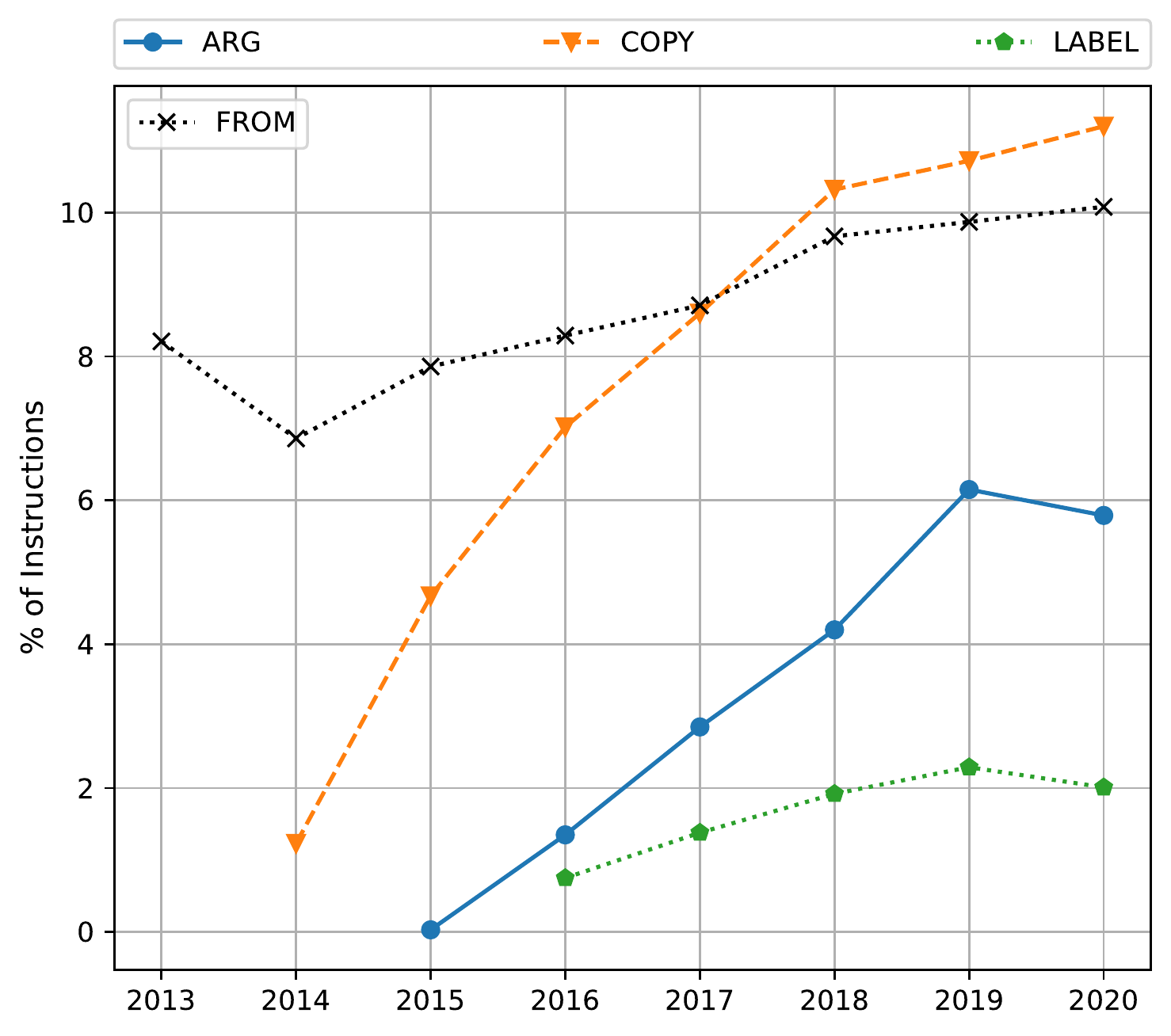}}
    \hfill
  \subfloat[New instructions less than 1\%\label{subfig:lt1}]{%
       \includegraphics[width=0.5\linewidth]{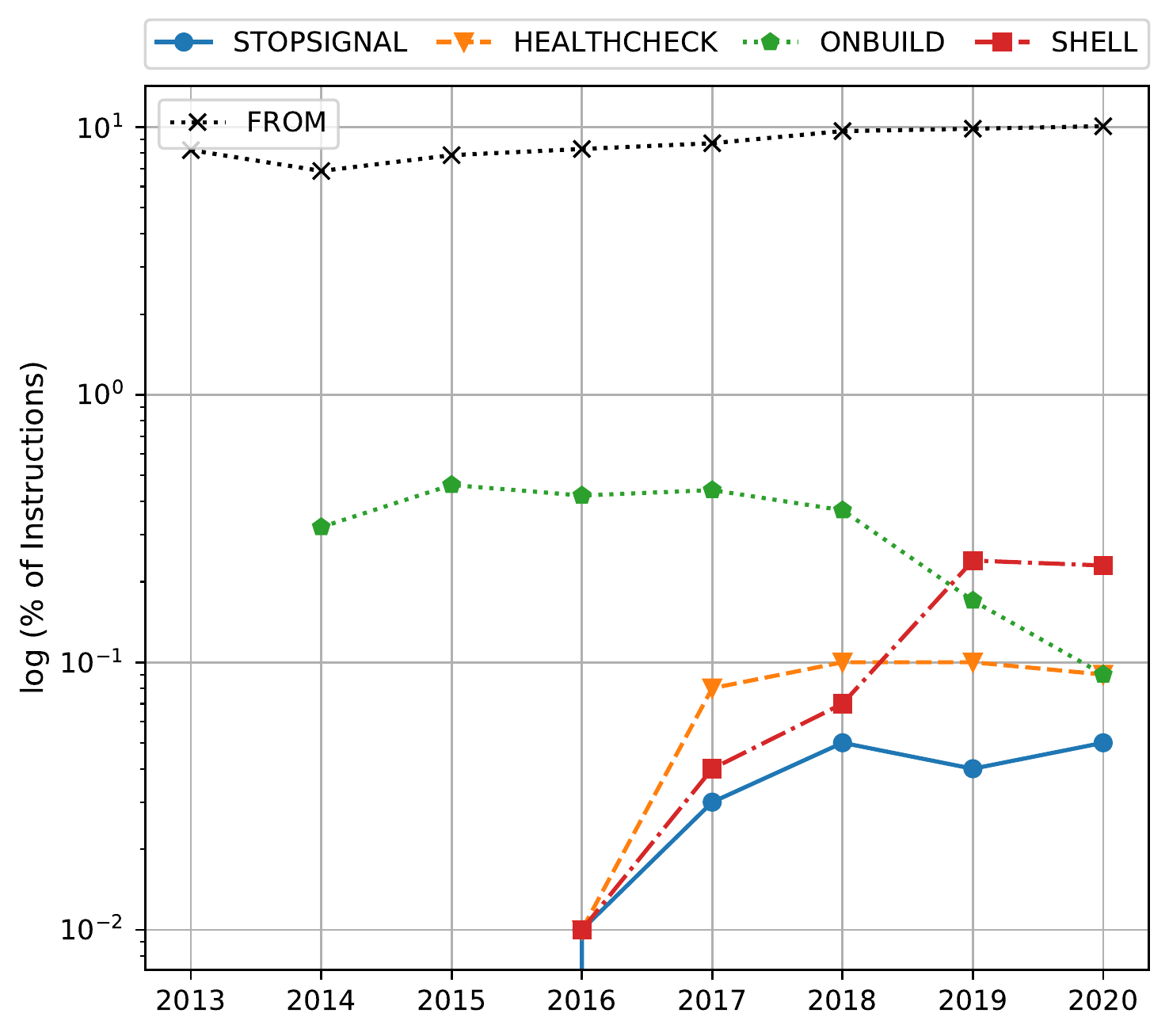}}
  \caption{Percentage of standard instructions ($\geq 2013$): (a) decreasing and (b) increasing, and new instructions ($\geq 2014$): (c) up to 12\% and (d) less than 1\%.}
  \label{fig:inc-dec}
\end{figure}

First, we test the dependence of Dockerfile instructions in 9,455,938 Dockerfiles and its usage per year by performing a chi-squared test and find that the variables are dependent on each other (p$<$0.001) which leads us to conclude that different amounts of instructions are used from 2013-2020. In the subsequent subsection of RQ1 we analyze the valid instructions of Dockerfiles using Figure~\ref{fig:inc-dec}. It should be noted that we perform filtering to remove instructions that are in years when they should not exist. These erroneous dates are due to the author creation timestamps of the commit which are created incorrectly and lead to uninterpretable results.



%
%

\subsection*{Instruction Usage}
We analyze the original 11 \qq{standard} instructions introduced with the release of Dockerfile in 2013 and the subsequent \qq{new} instructions released from 2014 onwards. We also attempt to find usage of \qq{INSERT} that was available for a short period until its deprecation in Docker version 0.4.5, but find no real usage of it. 

In Figure~\ref{fig:inc-dec} we present four charts broken down into two trends. In Figures~\cref{subfig:dec,subfig:inc} the standard instructions are split into two charts based on their increasing and decreasing trends. In Figures~\cref{subfig:gt10,subfig:lt1}, the new instructions are broken into two charts based upon the maximum peak percentage of instructions of all years so that increasing and decreasing trends can be visualized. 
To determine if an instruction appears more than once in a Dockerfile, we use the \qq{FROM} instruction as a baseline since it is present in every file --- if the percentage is greater than \qq{FROM}, then it likely appears more than once in a file and vice versa. Since the percentage of instructions can have large differences, we present the percentage of instructions in the log scale for Figures~\cref{subfig:dec,subfig:inc,subfig:lt1} to better visualize trends.

First, we can observe in Figure~\ref{subfig:dec} that all the instruction percentages decrease by 2020. However, we should note that the \qq{RUN} instruction begins at 58.01\% in 2013 and drops down to 37.27\% in 2020 --- a 20.74\% decrease that is larger than other instructions. Similarly, we also see \qq{ADD} drop 10.46\% from 12.32\% to 1.86\%. We hypothesize that the reason for the decrease in \qq{RUN} instructions is because commands can be combined into multi-line commands, some common commands are replaced with instructions like \qq{WORKDIR}, \qq{ENV}, and \qq{USER} which we can see increase in Figure~\ref{subfig:inc}.

\begin{figure}[tbp]
\centering

\includegraphics[width=0.75\linewidth]{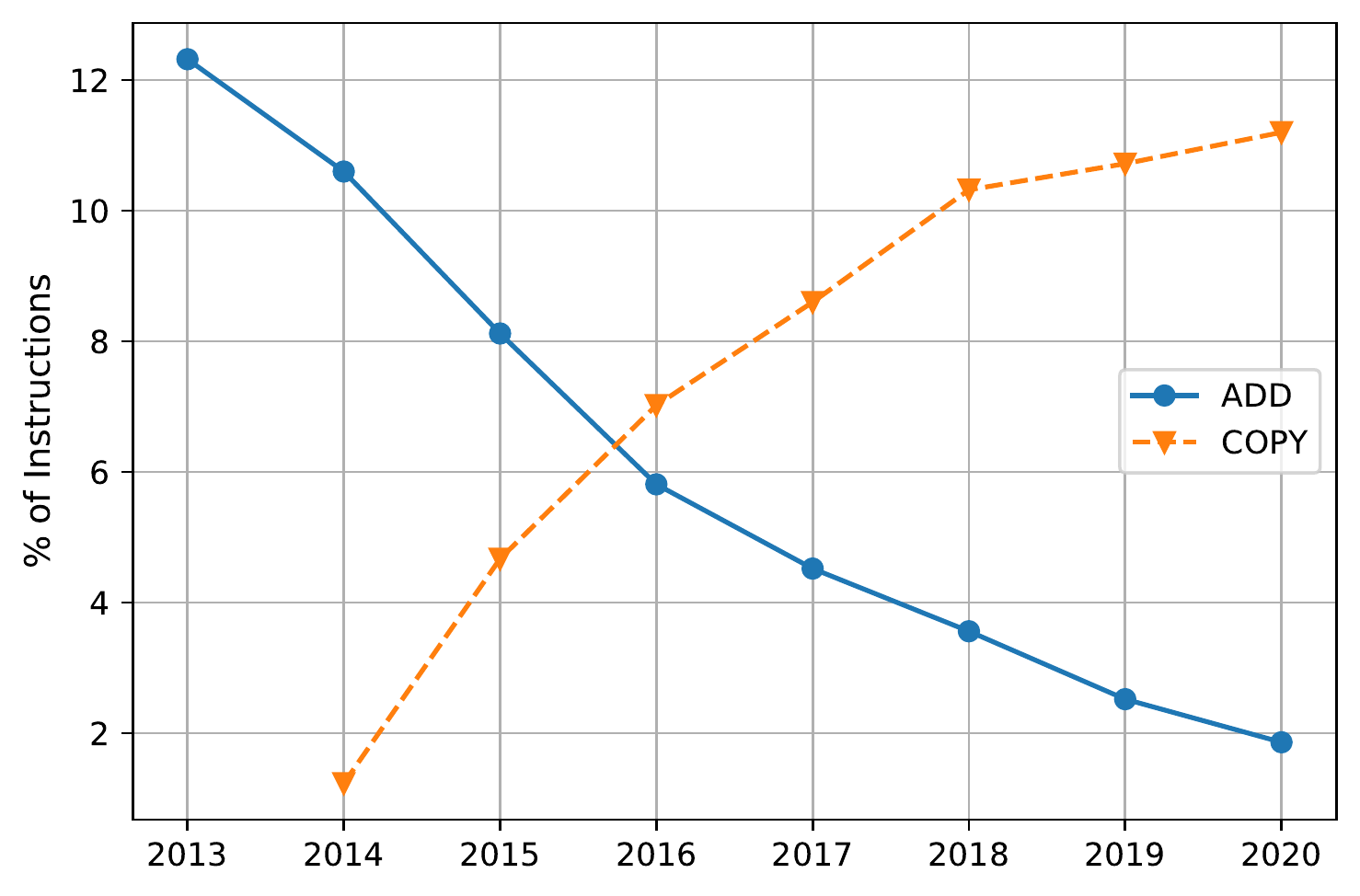}

\caption{The decrease of \qq{ADD} usage and increase of \qq{COPY} usage.}
\label{fig:cross-over-inst}
\squeezeup
\squeezeup
\end{figure}

In regards to the decrease of \qq{ADD} instructions, it is likely due to the introduction of the \qq{COPY} instruction which behaves similar to \qq{ADD} but does not do any post processing to a file once added to a container. To better visualize this relationship, we plot in Figure~\ref{fig:cross-over-inst} the percentage of \qq{ADD} and \qq{COPY} instructions over time where we can clearly see after the introduction of \qq{COPY} in 2014, it surpasses \qq{ADD} usage by 2016.

In Figure~\ref{subfig:inc}, we see that all instruction usage increases with \qq{ENV}, \qq{FROM}, and \qq{WORKDIR} appearing to visually increase at a faster rate than the other instructions. This is likely due to the fact that \qq{USER}, \qq{ENTRYPOINT}, and \qq{CMD} often appear at the end of Dockerfiles and are only used once. Notably, \qq{USER} appears to slightly increase which indicates that more containers may be started as a non-root user. The increase of \qq{FROM} is also interesting to note as in 2017, Docker began to support multi-stage builds which allows for multiple \qq{FROM} statements in a file. From 2017-2018  we see a sharper slope for \qq{FROM} compared to its subsequent years.

If we look at the new instructions in Figures~\cref{subfig:gt10,subfig:lt1}, we can see that almost all instructions are increasing in usage except for \qq{ONBUILD}. This is likely due to a decrease in usage of the \qq{ONBUILD} instruction as it is used specifically for Dockerfiles that will be a base image for another Dockerfile. Therefore, this trend suggests that many Dockerfiles may not be used as base images and those that are base images, do not employ the \qq{ONBUILD} instruction. Also of note is that the proportion of these instructions present throughout all Dockerfiles remains small as their sum of their counts account for less than 1\% of all instructions. This suggests that these instructions are for more niche use cases and possibly for more advanced users.



Our results of instruction usage cannot be compared directly to Cito et al.~\cite{cito2017empirical} as we do not count comment lines. Nonetheless, we note that \qq{RUN} is the top used instruction in our analysis similar to their previous findings. To better understand how \qq{RUN} instructions operate, we classify the \qq{RUN} instructions into 4 categories~\cite{cito_classify} in Table~\ref{tab:run-bd} using the preexisting classifiers of~\cite{cito2017empirical}. Additionally, we manually classify another 35 commands not present in the top 100 RUN instructions of the 9,455,938 Dockerfiles.

\begin{table}[tbp]
\centering
\caption{Percentage breakdown of \qq{RUN} instructions into 6 categories.}
\label{tab:run-bd}
\resizebox{\linewidth}{!}{%
\begin{tabular}{l|llllll}
\hline
               & \textbf{Dependencies} & \textbf{Filesystem} & \textbf{Permissions}  & \textbf{Build/Execute} & \textbf{Environment}            & \textbf{Other} \\ \hline
Examples                & apt-get, yum, npm    & mkdir, rm, cd       & chmod, chown, useradd & Rscript, node, php     & locale-gen, su, export &       \\ \hline
\textbf{RUN \%} & 44.94                 & 24.44               & 6.89                  & 6.49                   & 6.27                   & 10.96 \\ \hline
\end{tabular}%
}
\squeezeup
 \vspace{-2mm}
\end{table}

The classification results of the \qq{RUN} instructions appear to have a similar order to Cito et al.~\cite{cito2017empirical} (2016) and we find that dependency related commands are still the most used with \qq{RUN}. 


\subsection*{Common Instruction Groups}
Seeing that certain instructions are used more often than others, we wish to determine the most used groups of instructions among Dockerfiles. In Table~\ref{tab:inst-groups}, we breakdown the percentages of the top 10 instruction groups overall from 2013-2020.

\begin{table*}[tbp]
\centering
\caption{Overall top 10 instruction groups in percentages by year (top group for the year underlined).}
\label{tab:inst-groups}
\scriptsize
\begin{tabular}{llrrrrrrr}
\hline
\textbf{Instruction Grouping}              & \textbf{2013}       & \textbf{2014}       & \textbf{2015}       & \textbf{2016}       & \textbf{2017}       & \textbf{2018}       & \textbf{2019}        & \textbf{2020}        \\ \hline
CMD, COPY, ENV, EXPOSE, FROM, RUN, WORKDIR &      --               & 0.11                & 0.46                & 0.73                & 0.89                & 1.12                & 1.16                 & 1.16                 \\
CMD, COPY, ENV, FROM, RUN, WORKDIR         &    --             & 0.04                & 0.28                & 0.91                & 0.99                & 1.29                & 1.17                 & 1.32                 \\
CMD, COPY, EXPOSE, FROM, RUN, WORKDIR      &    --                 & 0.08                & 0.61                & 1.13                & 1.94                & 2.36                & 2.72                 & 3.03                 \\
CMD, COPY, FROM, RUN, WORKDIR              &     --                & 0.06                & 0.42                & 0.81                & 1.25                & 2.07                & 2.59                 & 2.91                 \\
CMD, ENV, FROM, RUN, WORKDIR               & 0.07                & 0.16                & 0.25                & 0.23                & 0.27                & 1.04                & 2.89                 & 3.82                 \\
CMD, ENV, FROM, RUN                        & 0.11                & 0.48                & 1.83                & 2.78                & 2.12                & \underline{\textbf{5.06}} & \underline{\textbf{11.03}} & \underline{\textbf{14.99}} \\
COPY, FROM, RUN, WORKDIR                   &     --                & 0.05                & 0.25                & 0.45                & 0.76                & 1.27                & 1.51                 & 1.59                 \\
COPY, FROM, RUN                            &     --                & 0.14                & 0.61                & 0.92                & 1.10                & 1.20                & 1.14                 & 1.12                 \\
ENV, FROM, RUN                             & 1.44                & 1.07                & 1.25                & 1.42                & 1.48                & 1.47                & 1.77                 & 1.10                 \\
FROM, MAINTAINER, RUN                      & \underline{\textbf{7.06}} & \underline{\textbf{5.09}} & \underline{\textbf{3.08}} & 2.47                & 1.55                & 0.86                & 0.39                 & 0.84                 \\
FROM, MAINTAINER                           & 0.15                & 0.31                & 0.29                & 0.34                & 0.54                & 0.72                & 1.45                 & 2.21                 \\
FROM, RUN                                  & 3.98                & 2.62                & 2.60                & \underline{\textbf{3.06}} & \underline{\textbf{3.39}} & 3.17                & 2.48                 & 2.85                 \\
FROM                                       & 0.18                & 0.33                & 0.42                & 0.48                & 0.68                & 2.26                & 1.17                 & 0.50                 \\ \hline
\end{tabular}
\squeezeup
\end{table*}

From the table, we can observe that \qq{FROM} is in every group and \qq{RUN} is almost in every group. What is interesting to note is the trend of the most popular groups (highlighted by the bold and underlined cells in the table). From 2013-2015, the most popular instruction groups of the years contain \qq{MAINTAINER}. But from 2016 onwards, \qq{FROM} and \qq{RUN} without \qq{MAINTAINER} become more popular. Notably, \qq{MAINTAINER} becomes deprecated in 2017.

We also notice that from 2018 onwards, the most popular group contains \qq{CMD}. What is particularly interesting to note is that Dockerfiles should contain either a \qq{CMD} or \qq{ENTRYPOINT} instruction so that no additional flags need to be specified when running the container~\cite{cmd-entry}. Therefore, it appears that more Dockerfiles from 2018 onwards should be able to run as more Dockerfiles contain \qq{CMD} and therefore no command needs to be passed to the image when starting the container.
Also of note is the larger percentage increases for \qq{CMD, ENV, FROM, RUN} from 2018-2020. This may be due to the fact these are the minimal amount of commands needed to run software using a base image: \qq{FROM} is used to define a base image, \qq{ENV} is used to set the relevant environment variables, \qq{RUN} is used to run a set of commands for setting up the container, and \qq{CMD} is used to setting a default command to run for the container.

Overall, we see \qq{RUN} in the top groups of instructions in Table~\ref{tab:inst-groups} with the size of the group of instructions increasing from 2018-2020. This leads credence to our hypothesis that \qq{RUN} command usage may have decreased overall because of the increase of usage of other instructions in-conjunction with \qq{RUN}. This is also supported by our observation that the amount of instructions in Dockerfiles over the years remain relatively constant.

\subsection*{Base Image Usage}
\begin{figure}[tbp] 
    \centering
  \subfloat[Top 20 base images \label{subfig:top20_all}]{%
       \includegraphics[width=0.5\linewidth]{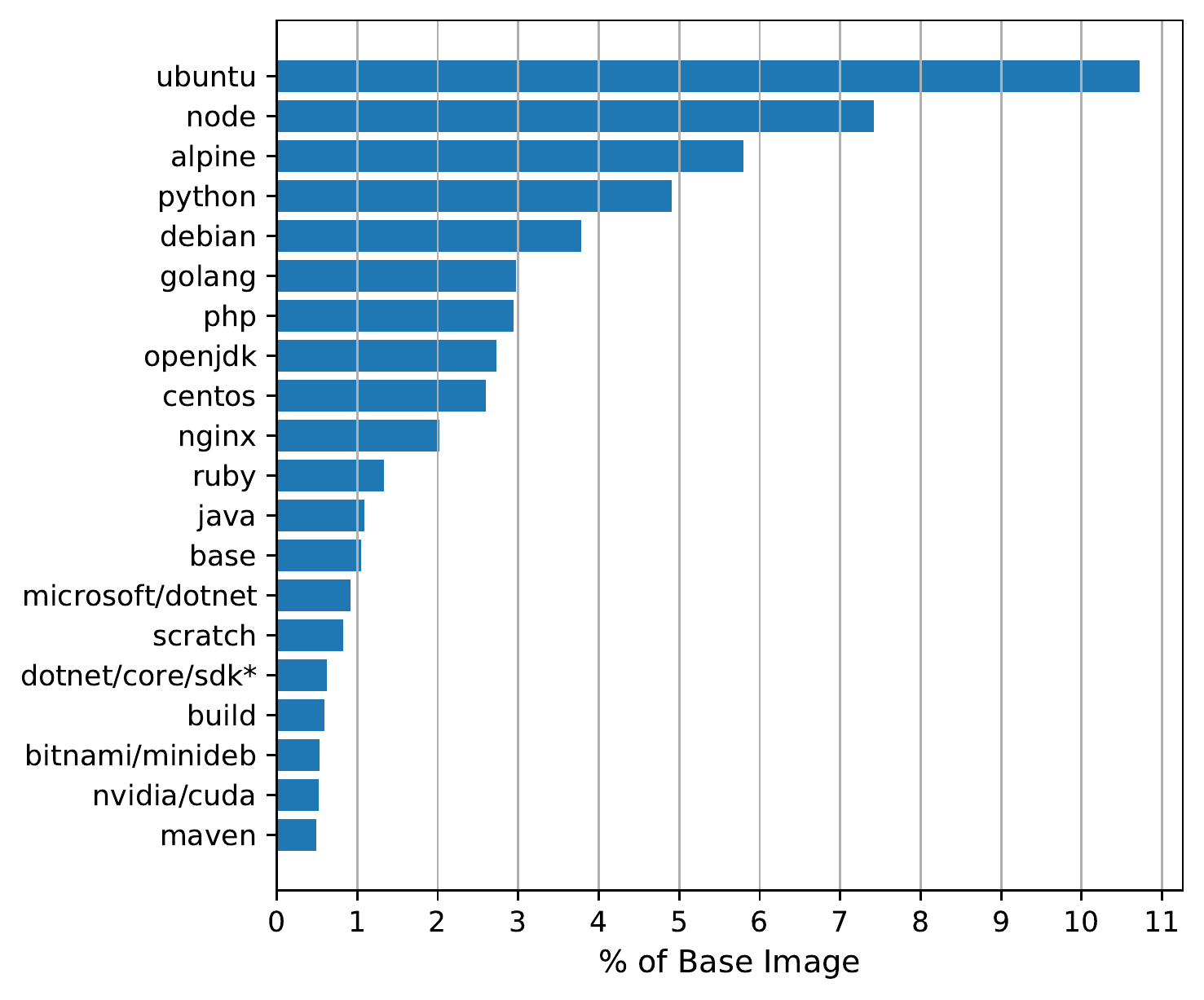}}
    \hfill
  \subfloat[Top 20 base images categorized\label{subfig:top20_cat}]{%
        \includegraphics[width=0.5\linewidth]{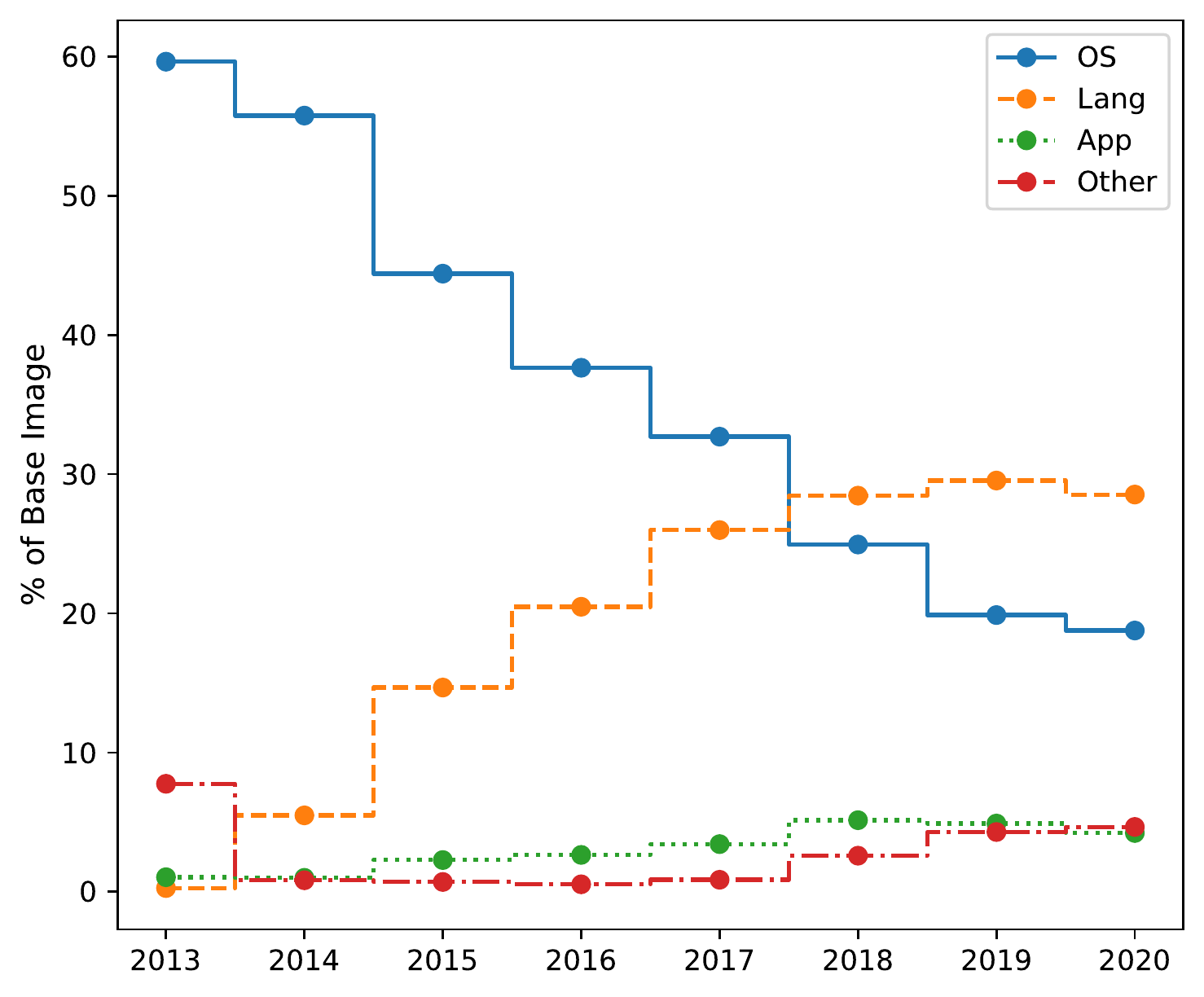}}
        \\
  \subfloat[Application/Language images of Top 10\label{subfig:top10_lang}]{%
       \includegraphics[width=0.5\linewidth]{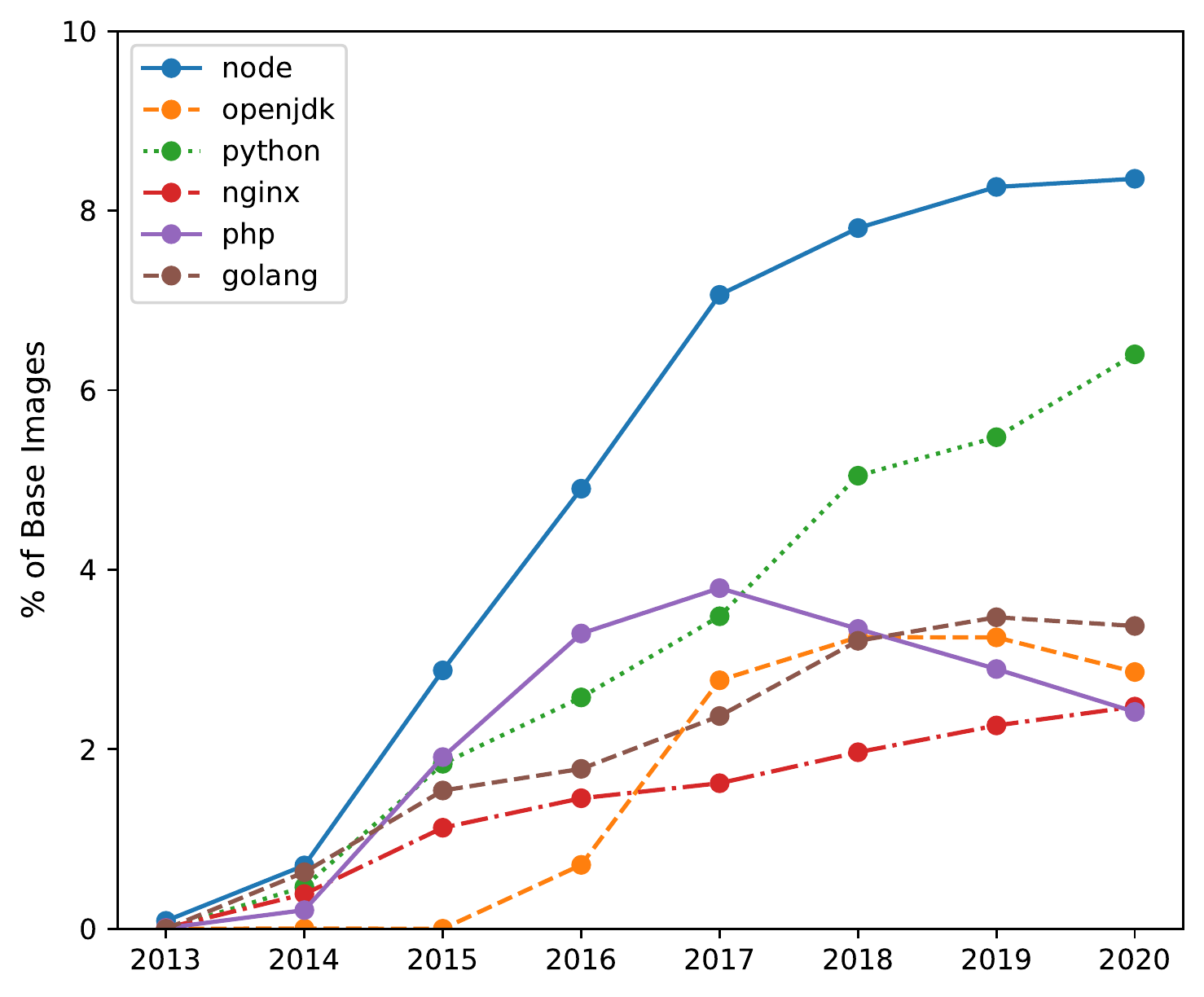}}
    \hfill
  \subfloat[OS Images of Top 10\label{subfig:top10_OS}]{%
        \includegraphics[width=0.495\linewidth]{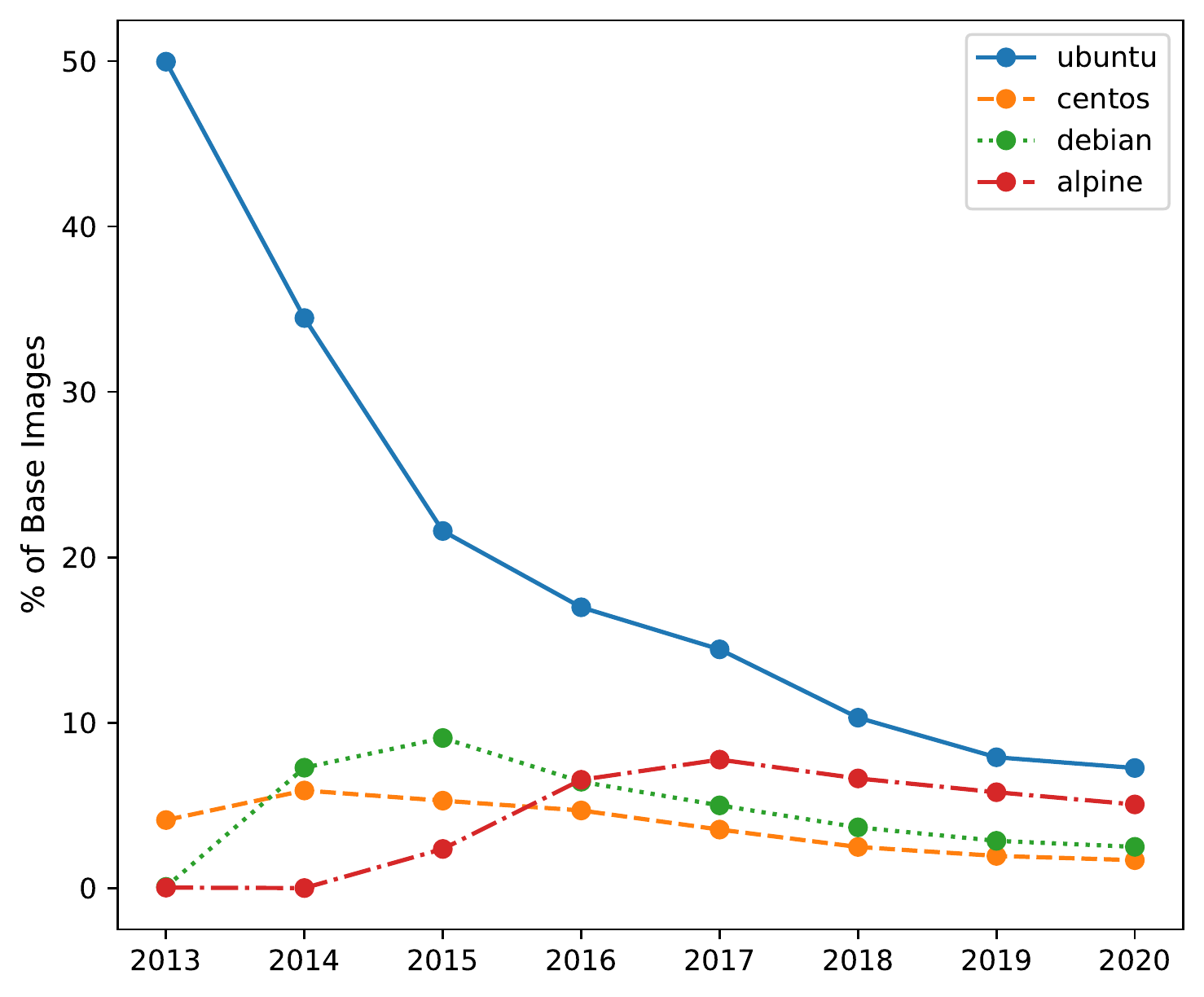}}
  \caption{Percentage of: (a) overall top 20 base images (Note that \qq{dotnet/core/sdk*} is \qq{mcr.microsoft.com/dotnet/core/sdk}) (b) overall top 20 base images categorized per year, and top 10 base images: (c) broken down by application/language and (d) broken down by OS.}
  \label{fig:base-image} 
  \squeezeup
  \squeezeup
\end{figure}

As base images are part of the \qq{FROM} instruction (mandatory in every Dockerfile) and the basis for creating a Docker image, we investigate what are some of the most common occurring base images in Dockerfiles. In Figure~\ref{fig:base-image}, we present the top 20 base images overall which are then categorized and broken down into their usage over their years. Furthermore, we also isolate the application/language and OS images of the top 10 base images and plot them into two charts. 

First, we can see that Ubuntu is the most popular base image among the top 20 overall in Figure~\ref{subfig:top20_all} similar to the previous findings of Cito et al.~\cite{cito2017empirical} (2016). However, we note that Ubuntu's popularity is trending downwards like Lin et al.~\cite{lin2020large} (2020). To better characterize the base images we categorize them into four categories: operating system images like Ubuntu and Debian as \qq{OS}, programming image languages like node and python as \qq{Lang}, application image languages like nginx as \qq{App}, and other images that are base images like base, build, scratch as \qq{Other}. In total, we classify 55 different kinds of base images into four categories.

We plot the categorizations in Figure~\ref{subfig:top20_cat}, and perform a KW test on the distributions noting significance with a subsequent MW rank test finding that base image categories \qq{App} and \qq{Lang} are insignificant with p=0.28 meaning that trends between application and language images are similar as opposed to trends between language and OS images being significantly different. As such, the significance of language images exceeding OS images in 2018 confirms the results of Lin et al.~\cite{lin2020large} one year earlier, it also demonstrates that the trend of base image usage follows the recommendations of Cito et al.~\cite{cito2017empirical} (2016) to not use OS images as base images since they can make Docker images larger. Interestingly, we also note that \qq{Other} has a higher percentage in 2013 which we attribute to Dockerfiles initially specifying \qq{base} as part of the \qq{FROM} instruction as evidenced in some 2013 online articles~\cite{atlassian,eclipsesource}.

Next, we observe the usage of different application/language images in Figure~\ref{subfig:top10_lang} and OS images in Figure~\ref{subfig:top10_OS}. Most notably, there are no application/language images in 2013 and its usage only began in 2014. This is likely due to Docker Hub being first announced in 2014 to simplify the process for the distribution of container images~\cite{dockerhub_announce}. This suggests that Docker Hub might have been a key role in proliferating the use of different base images in Dockerfiles.
Another observation that we should note is that OS images appear to have high usage in the beginning likely due to non-optimized OS initially being suggested as the base images to run applications as evidenced in the documentation examples~\cite{docker_orig_doc}. However, based on the downward trend of OS images, we hypothesize that as the ecosystem matures, images become more domain specific, efficient, and lightweight.

 \noindent \hrulefill
 
  \noindent \textit{\textbf{Dockerfile Instructions and Base Images:}} As the Docker ecosystem matures, its usage appears to be more streamlined. More lightweight images are being used as opposed to OS images. Furthermore, Dockerfiles appear to use a more diverse set of instructions.
 
 \noindent \hrulefill







\subsection*{{\normalfont\textbf{RQ2. How prevalent are code smells in Dockerfiles?}}}
Previous studies use the Haskell Dockerfile Linter (Hadolint)~\cite{hadolint} to detect if Dockerfiles follow best practices. In this section, we replicate~\cite{lin2020large,cito2017empirical,wu2020characterizing} with a larger dataset to see how Dockerfile smells may differ. Furthermore, we also consider when instructions are first introduced as described in Section~\ref{dockerfile-format} in order to remove smells that may not be applicable to a year. We outline what we remove in Table~\ref{tab:rule-removal}. It should also be noted that we are unable to detect DL3012 (Provide an email address or URL as maintainer) for earlier Dockerfiles that contain \qq{MAINTAINER} instructions due to being deprecated from Hadolint since 2017~\cite{dl3012}. 

\begin{table}[tbp]
\centering
\caption{Hadolint rules removed by year.}
\label{tab:rule-removal}
\resizebox{\linewidth}{!}{%
\begin{tabular}{cll}
\hline
\textbf{Removal Year}          & \textbf{Rule} & \textbf{Description}                                                     \\ \hline
All            & DL3026        & Use only an allowed registry in the FROM image                           \\
\textless 2014 & DL3020        & Use COPY instead of ADD for files and folders.                           \\
\textless 2014 & DL3021        & COPY with more than 2 arguments requires the last argument to end with / \\
\textless 2014 & DL3022        & COPY --from should reference a previously defined FROM alias             \\
\textless 2014 & DL3023        & COPY --from cannot reference its own FROM alias                          \\
\textless 2016 & DL4005        & Use SHELL to change the default shell                                    \\
\textless 2016 & DL4006        & Set the SHELL option -o pipefail before RUN with a pipe in it            \\
\textless 2017 & DL4000        & MAINTAINER is deprecated                                                 \\ \hline
\end{tabular}%
}
\squeezeup
\end{table}

\begin{figure}[tbp]
\centering

\includegraphics[width=0.8\linewidth]{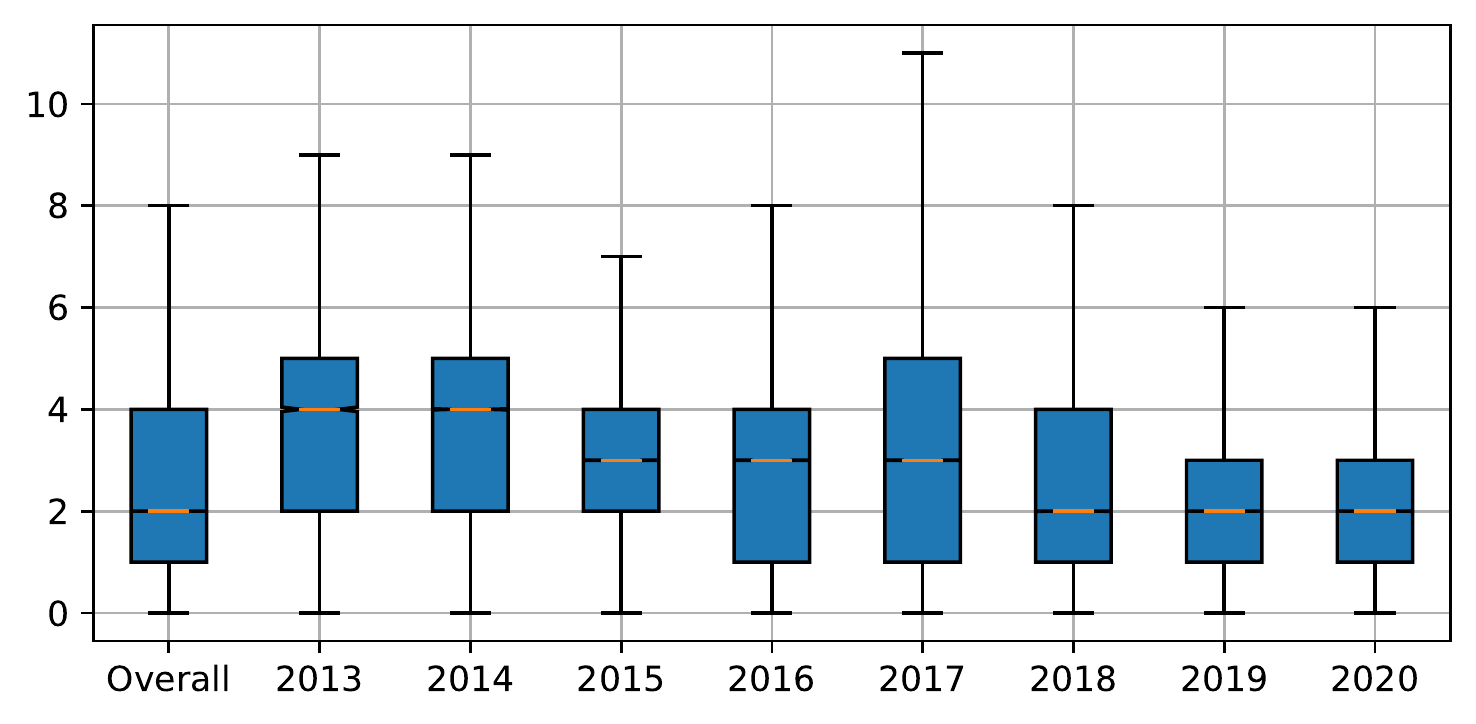}

\caption{Number of smells from 2013-2020.}
\label{fig:smells-time}
\squeezeup
\end{figure}

We begin by observing the overall Dockerfile smell count over the years in Figure~\ref{fig:smells-time} for 9,445,029 blobs. When a KW test is performed on the years 2013-2020, it is found to be significant (p$<$0.001) and with a subsequent MW rank test all pairs are significant (p $\leq$6.94e-51).
Therefore, we can say that overall in the latter years of 2019-2020 Dockerfile smells have decreased. This decreasing trend of Dockerfile smells is similar to the observations of Lin et al.~\cite{lin2020large} (2020) and also Wu et al.~\cite{wu2020characterizing} (2020) where it is found that newer projects have fewer smells.

\begin{figure}[tbp] 
    \centering
  \subfloat[Increasing Dockerfile smells\label{subfig:df-dec}]{%
       \includegraphics[width=0.5\linewidth]{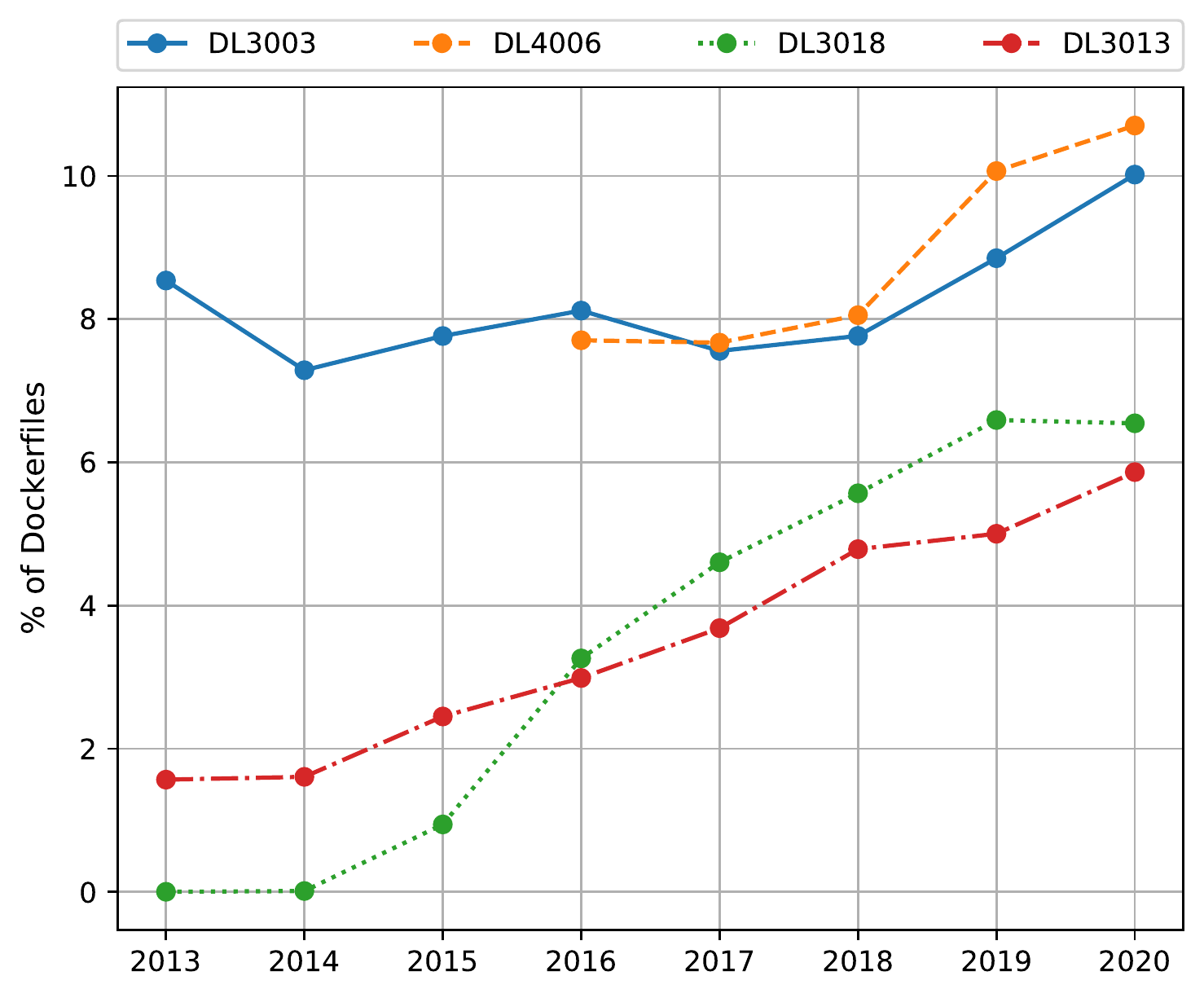}}
    \hfill
  \subfloat[Decreasing Dockerfile smells\label{subfig:df-inc}]{%
        \includegraphics[width=0.5\linewidth]{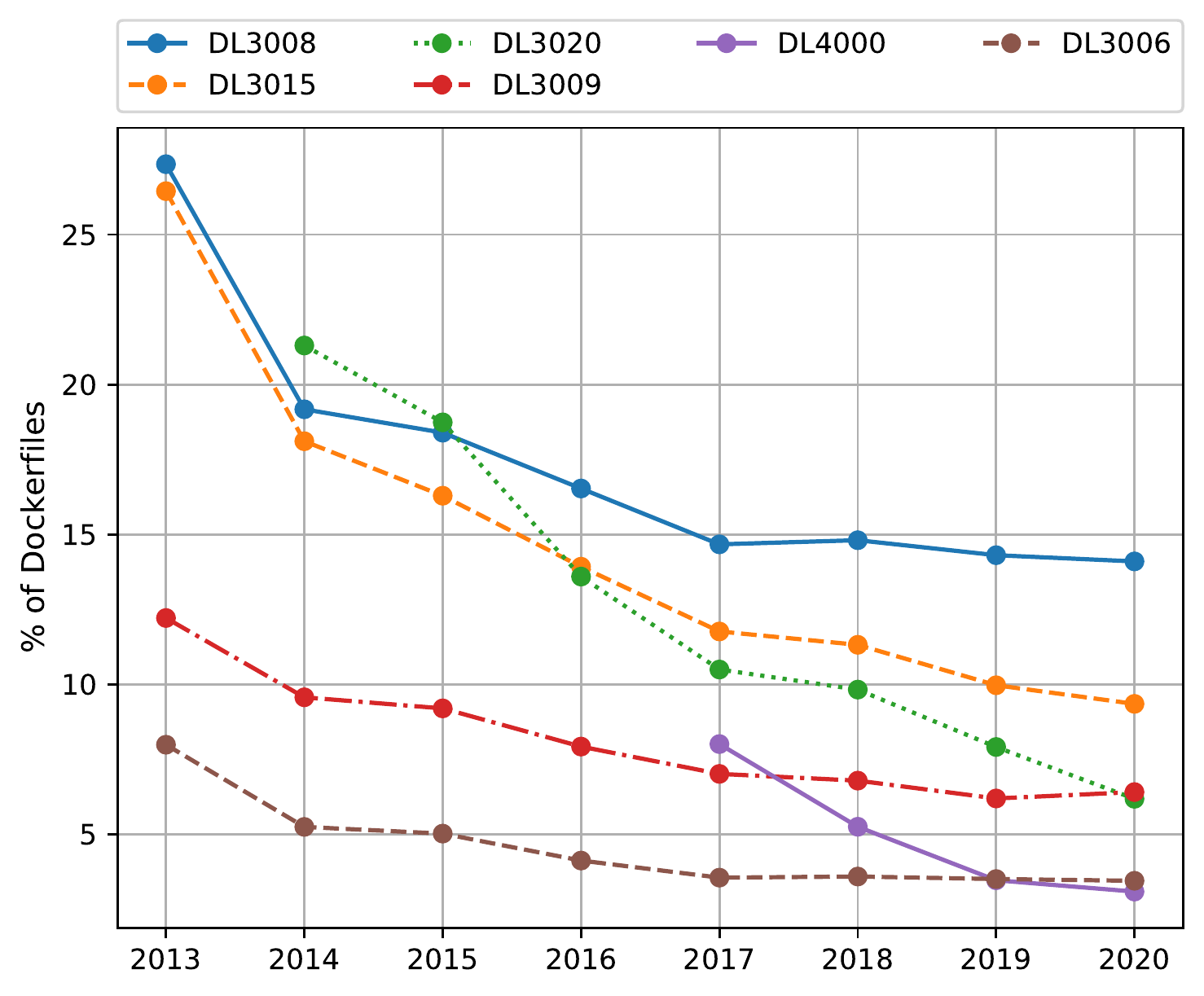}}
  \caption{Percentage of top 10 Dockerfile smells: (a) increasing and (b) decreasing.}
  \label{fig:df-inc-dec} 
  \squeezeup
  \squeezeup
\end{figure}

Next, we drill down and observe the top 10 overall Dockerfile smells found in Table~\ref{tab:top10df-smells} split into decreasing and increasing trends as seen in Figure~\ref{fig:df-inc-dec}.

\begin{table}[tbp]
\centering
\caption{Top 10 Overall Dockerfile Smells.}
\label{tab:top10df-smells}
\resizebox{\linewidth}{!}{%
\begin{tabular}{lll}
\hline
\textbf{Rule} & \textbf{Percentage} & \textbf{Description}                                            \\ \hline
DL3008        & 15.08               & Pin versions in apt get install                                 \\
DL3015        & 11.50               & Avoid additional packages by specifying --no-install-recommends \\
DL3020        & 10.05               & Use COPY instead of ADD for files and folders                   \\
DL3003        & 8.43                & Use WORKDIR to switch to a directory                            \\
DL4006        & 8.20                & Set the SHELL option -o pipefail before RUN with a pipe in      \\
DL3009        & 6.97                & Delete the apt-get lists after installing something             \\
DL3018        & 5.16                & Pin versions in apk add                                         \\
DL3013        & 4.43                & Pin versions in pip                                             \\
DL4000        & 3.84                & MAINTAINER is deprecated                                        \\
DL3006        & 3.74                & Always tag the version of an image explicitly                   \\ \hline
\end{tabular}%
}
\squeezeup
\end{table}

\subsection*{Dockerfile Smell Trends}
Most Dockerfile rules violated are similar to Lin et al.~\cite{lin2020large} (2020) except for DL3013 and DL3018 which replaces the smells of DL3025 (Not using JSON notation for CMD) and DL3007 (Using the error-prone latest tag) in~\cite{lin2020large}. It is interesting to note that version pinning is still a major smell that appears similar to findings to Cito et al.~\cite{cito2017empirical} (2016) and Lin et al.~\cite{lin2020large}. Unlike previous findings, however, we note that version pinning smells are trending upwards. 

Dockerfile smells such as using \qq{COPY} instead of \qq{ADD} appear to be trending downwards which also follows our observed trend of increased \qq{COPY} usage in Figure~\ref{fig:cross-over-inst}. It is also interesting to note that DL4000 is quickly reduced which suggests that Dockerfiles have removed the \qq{MAINTAINER} instruction quite quickly.

Of note is DL4006 increasing as it appears to be an obscure smell that would not be immediately noticeable to a typical end user that might use piping in their \qq{RUN} command. Therefore, better user education might be needed to raise awareness to this issue. The increase of DL3003, not using \qq{WORKDIR} to change directories, also suggests that better education of available instructions might also be needed.

 \noindent \hrulefill
 
 \noindent \textit{\textbf{Dockerfile Smells:}} Dockerfile smells appear to be trending downwards overall. However, version pinning to reduce dependency issues still appears to be a major issue that is increasing over the years. It is also interesting to note that the smell of using \qq{COPY} instead of \qq{ADD} is trending downwards which we also notice follows the trend of increased \qq{COPY} usage instead of \qq{ADD}.
 
 \noindent \hrulefill

\subsection*{{\normalfont\textbf{RQ3. How do Dockerfiles change over time?}}}
\begin{figure}[tbp]
\centering
\includegraphics[width=0.8\linewidth]{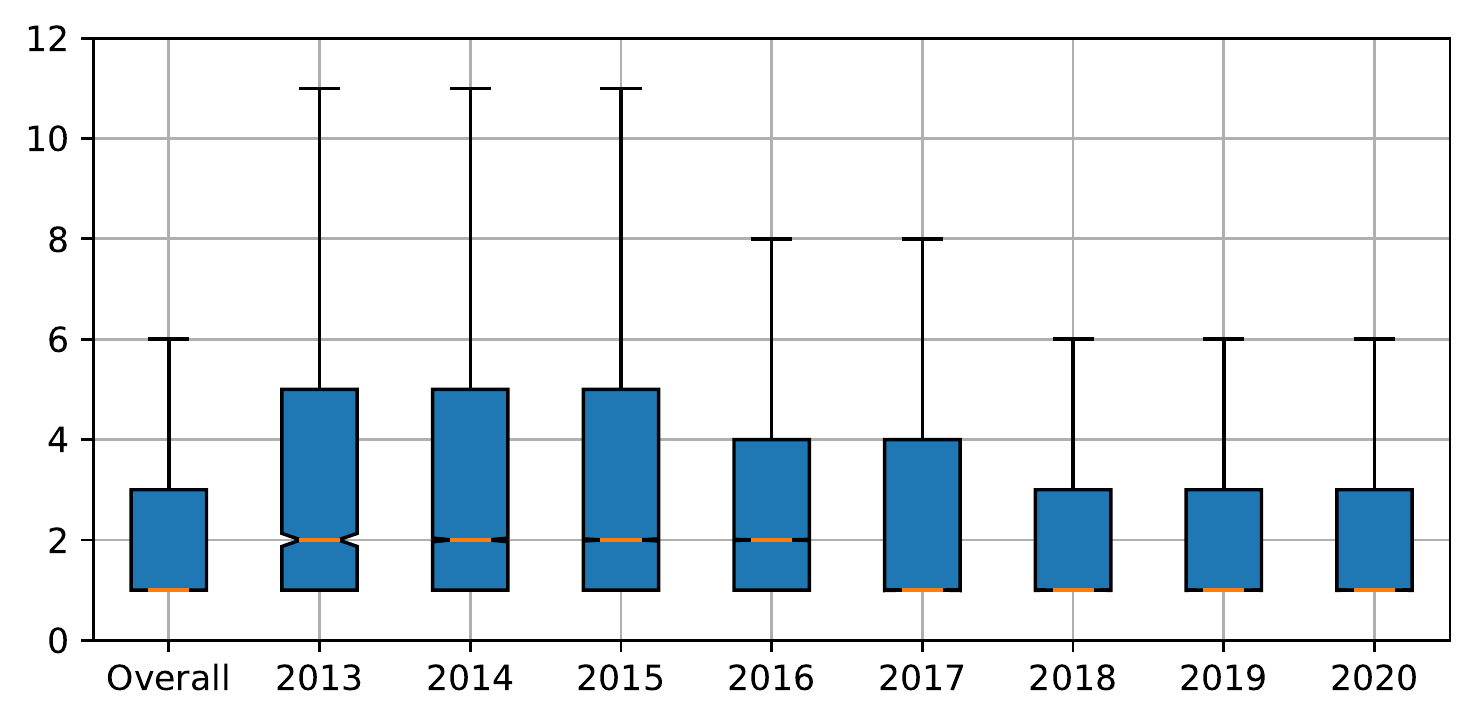}
\caption{Count of revisions containing Dockerfiles from 2013-2020.}
\label{fig-rev-year}
\squeezeup
\squeezeup
\end{figure}

To understand how often Dockerfiles are changed, we present in Figure~\ref{fig-rev-year} the count of commits containing Dockerfiles grouped by year. We count each commit as a revision with each repository having at least 1 revision. In the box plots trending over the years, we see that overall there are only 3 revisions per year, and 50 percent of revisions have only 1 revision. However, there is great variation with a standard deviation of 149 revisions. We perform KW test finding significance (p$<$0.001) and MW rank test finding only the pair 2013 and 2014 insignificant (p=0.99). Thus, we can also say that there appears to be a lower amount of revisions in the later years which may be due to having more ready-to-use Docker images and therefore not needing custom Dockerfiles.

We also include the overall distribution of commit counts per year with a standard deviation of 149 and find up to 75\% as having 3 revisions and up to the median of 1 which means there are no changes within a year. The revisions per year appear low, similar to findings by Cito et al.~\cite{cito2017empirical} (2016).

In the subsequent subsections, we analyze the magnitude and nature of change in Dockerfiles like Cito et al.~\cite{cito2017empirical}. For the data, we retrieve the earliest Dockerfile commit for each distinct repository with more than one commit and their related Dockerfile blobs resulting in 1,765,277 unique SHA-1s. To ensure that the blobs have any lineage, we query the \qq{ob2b} mapping in WoC to determine if any children exist and find 1,121,849 pairs. We also determine if the parent and children blob contents can be retrieved, finding the number of blob pairs to be 1,115,337. To further reduce the pairs, we also check to see if the parents are children and remove those blobs resulting in 1,036,546 blob pairs. Noting that there are multiple pathways from parent to children and wishing to see a more diverse set of evolving blobs, we select each parent only once resulting in 848,051 blob pairs.

Due to the computational intensity of calculating the history from 848,051 blob pairs, we randomly sample without replacement 16,321 blob pairs using the random seed 69780. The sample size is determined using Cochran's formula~\cite{cochran1963sampling} as described by Israel~\cite{israel1992determining} with a Z-Score of 2.58, 99\% confidence level, maximum variability of 0.5, and a precision of 1\%.


To trace the lineage of the blobs, we begin with the 16,321 blobs and query a single path found with the WoC \qq{ob2b} mapping calculating the lines added and deleted with the \textit{diffstat}~\cite{diffstat} and noting the lines changed with \textit{diff}~\cite{diff}. It should be noted that we only get the insertion and deletion difference as we use the unified format of diff. From our initial starting blob pairs, we observe the differences among 26,858 pairs of blobs and 42,923 distinct SHA-1s.

\subsection*{Magnitude of Change}
\begin{figure}[tbp]
\centering
\includegraphics[width=0.75\linewidth]{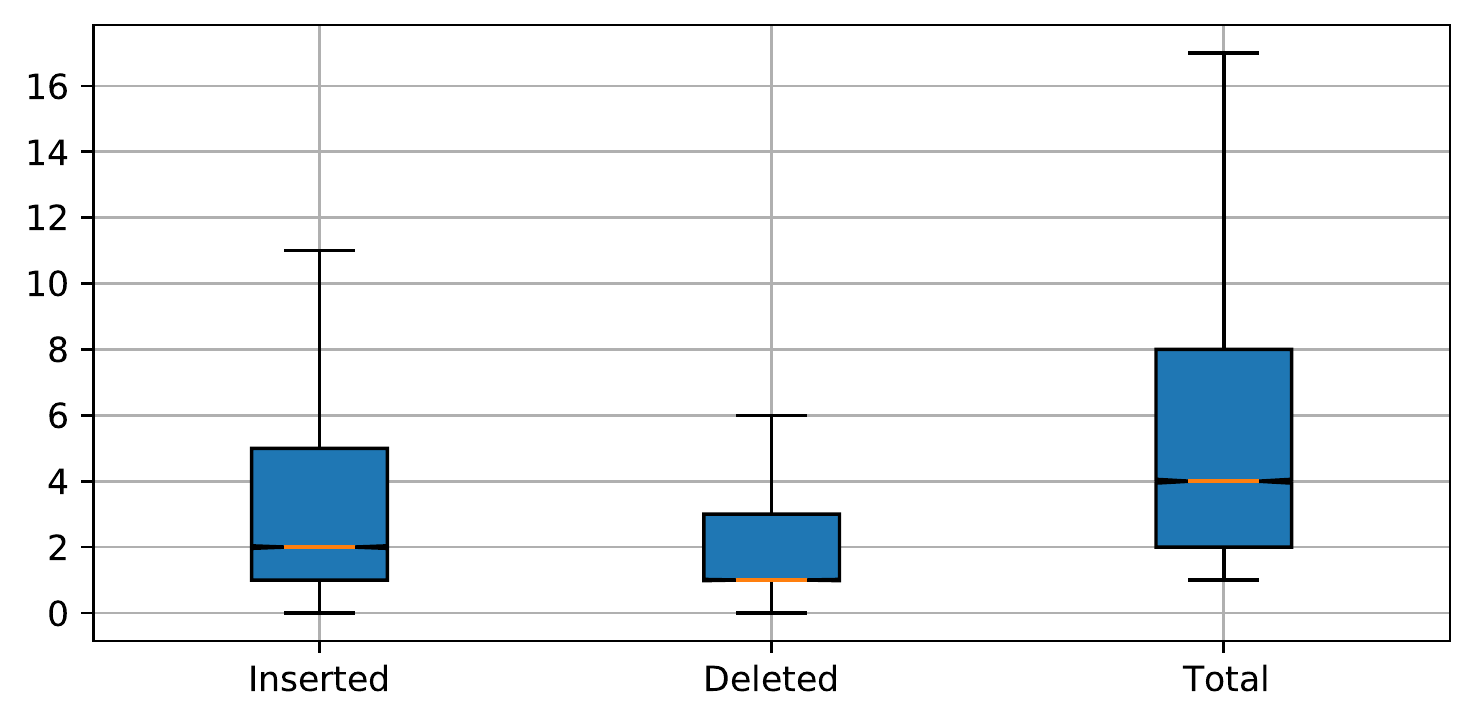}
\caption{Number of lines inserted, deleted, and the total count.}
\label{fig-mag-change}
\squeezeup
\squeezeup
\end{figure}

To understand the magnitude of change, we refer to Figure~\ref{fig-mag-change} which contains box plots for lines inserted, lines deleted, and the total of lines inserted and deleted. We perform a KW test between the lines inserted and deleted and find that they are statistically significant (p$<$0.001). Therefore, we can say for our sampled changes of Dockerfiles that there are more line insertions than deletions overall like Cito et al.~\cite{cito2017empirical} (2016).

With insertions, there is a mean of 5.14 and standard deviation of 11.08 with up to 75\% of changes having 5 lines inserted, up to 50\% having 2 lines inserted, and up to 25\% having 1 line inserted. In comparison, deletions have a mean of 3.77 and standard deviation of 9.1 with up to 75\% of changes having 3 lines deleted and up to 50\% having 1 line deleted. We see that overall, there are not very many lines changed as the count of up to 75\% of total lines changed is 8 with a standard deviation of 18.18.

\vspace{-1.5mm}
\subsection*{Nature of Change}

Next, we look specifically look at the lines that are changed which include Dockerfile instructions. We break down the total Dockerfile instructions in Table~\ref{tab:add-del} in terms of insertions and deletions. We also perform KW test on the distribution of Dockerfile instructions and find them to be statistically insignificant with p=0.61, therefore we can say that there are no differences in the instructions that are inserted or deleted.

\begin{table}[tbp]
    \begin{subtable}[h]{0.45\linewidth}
        \centering
        \scriptsize
        \begin{tabular}{lr}
        \hline
        \textbf{Instruction} & \textbf{\%} \\ \hline
        RUN                  & 38.93       \\ \hline
        {\tiny Dependencies}                                              & {\tiny 48.24}                                     \\
        {\tiny Filesystem}                                               & {\tiny 22.84}                                     \\
        {\tiny Permissions}                                               & {\tiny 7.10}                                     \\
        {\tiny Build/Execute}                                               & {\tiny 6.38}                                     \\
        {\tiny Environment}                                               & {\tiny 3.00}                                     \\ \hline
        COPY                 & 12.60       \\
        FROM                 & 10.47       \\
        ENV                  & 10.12       \\
        CMD                  & 5.02        \\
        ADD                  & 4.50        \\
        WORKDIR              & 4.19        \\
        EXPOSE               & 3.22        \\
        ARG                  & 3.16        \\
        ENTRYPOINT           & 2.63        \\
        LABEL                & 1.40        \\
        MAINTAINER           & 1.23        \\
        USER                 & 1.13        \\
        VOLUME               & 1.07        \\
        ONBUILD              & 0.16        \\
        HEALTHCHECK          & 0.09        \\
        SHELL                & 0.05        \\
        STOPSIGNAL           & 0.03        \\ \hline\\
        \end{tabular}
       \caption{Instructions Inserted}
       \label{tab:add}
    \end{subtable}
    \hfill
    \begin{subtable}[h]{0.45\linewidth}
        \centering
        \scriptsize
        \begin{tabular}{lr}
        \hline
        \textbf{Instruction} & \textbf{\%} \\ \hline
        RUN                  & 38.44       \\ \hline
        {\tiny Dependencies}                                              & {\tiny 49.80}                                     \\
        {\tiny Filesystem}                                               & {\tiny 22.32}                                     \\
        {\tiny Permissions}                                               & {\tiny 6.21}                                     \\
        {\tiny Build/Execute}                                               & {\tiny 6.14}                                     \\
        {\tiny Environment}                                               & {\tiny 3.30}                                     \\ \hline
        FROM                 & 12.58       \\
        COPY                 & 11.01       \\
        ENV                  & 9.79        \\
        CMD                  & 5.93        \\
        ADD                  & 4.79        \\
        WORKDIR              & 4.09        \\
        EXPOSE               & 3.02        \\
        ENTRYPOINT           & 2.88        \\
        ARG                  & 2.48        \\
        MAINTAINER           & 1.70        \\
        LABEL                & 1.18        \\
        VOLUME               & 1.05        \\
        USER                 & 0.81        \\
        ONBUILD              & 0.15        \\
        HEALTHCHECK          & 0.06        \\
        SHELL                & 0.02        \\
        STOPSIGNAL           & 0.01        \\ \hline\\
        \end{tabular}
        \caption{Instructions Deleted}
        \label{tab:del}
     \end{subtable}
     \caption{Percentage of total Dockerfile instructions that are (a) inserted and (b) deleted with \qq{RUN} instruction broken down into categories.}
     \label{tab:add-del}
     \squeezeup
     \squeezeup
     \vspace{-2mm}
\end{table}

In Table~\ref{tab:add-del}, when the \qq{RUN} is broken down by the first command executed, we see that dependencies make up about half of all \qq{RUN} instructions. Interestingly, the other top instructions also deal with changing files or settings to run software such as \qq{COPY} to move new files, \qq{FROM} to change base images, and \qq{ENV} to set new environment variables. The breakdown of changes also appear similar to Cito et al.~\cite{cito2017empirical} (2016).


 \noindent \hrulefill
 
 \noindent \textit{\textbf{Dockerfile Revisions:}} Dockerfiles have a low amount of revisions per year and revisions appear to slowly trend downwards from 2013-2020. Within the changes, there are more insertions than deletions. The Dockerfile instructions that are changed or deleted often have to deal with running new commands, moving new files, changing the base image or setting new environment variables. Within the \qq{RUN} instruction, commands dealing with dependencies are the most common. 
 
  \noindent \hrulefill
\vspace{-1.5mm}



\section{Threats to Validity}
\label{sec:ttv}
With regards to construct validity, we assume that WoC data is accurately mined. Furthermore, we must consider that the data cleaning and alignment of the data obtained from WoC has been done correctly. To address these concerns, we provide a replication dataset containing source code files from obtaining, processing, and generating our results. Furthermore, when parsing Dockerfiles, we assume that they may have been correctly parsed. However, there may be some slight inaccuracies due to previous versions possibly being correctly parsed in previous Docker versions but not the latest version. 


For internal validity, we note that the blobs we analyze may not actually be used in the real world for software deployments and instead are used for instructional or toy projects. Therefore, we cannot assume that the hypotheses of our findings apply to all the blobs in our dataset. Nonetheless, the dataset we analyze is extremely large and therefore those blobs are likely insignificant. 
Additionally, when we sample the beginning blobs to get a representative sample of how Dockerfiles evolve for RQ3, we only consider the very first Dockerfile blob commit of a distinct repository. Some Dockerfiles may not have been part of the very first Dockerfile blob commit. In addition, our findings from the sample may produce results that are coincidental, but we use the Cochran formula with simple random sampling to ensure a representative sample is used.


Finally, we note the threats to external validity. WoC attempts to mine all possible OSS repositories, but is by no means comprehensive and therefore might not generalize to all projects.
Furthermore, the Dockerfiles are analyzed at a blob level as opposed to a project level and therefore the trends presented may also not generalize to data when analyzing on a per project basis. It should also be noted that only public repositories are analyzed, and private repositories may create Dockerfiles differently. Despite these shortcomings, this paper analyzes a dataset larger than any previous studies and appears to confirm many findings of previous studies.


\section{Conclusion}
\label{sec:con}
Using the largest and most recent (as of December 2020) Dockerfile dataset known to date, consisting of over 9.4 million unique Dockerfiles between 2013-2020, we find similar trends on a year by year basis confirming observations of previous studies~\cite{cito2017empirical,lin2020large,wu2020characterizing, henkel2020learning}. As such, we provide suggestions to developers who use Docker and write Dockerfiles, and outline a benefit of our study for the maintainers of Docker.

To support developers, we echo Cito et al.'s~\cite{cito2017empirical} (2016) recommendations to not use heavy OS images which creates large footprints for containers and to use version pinning for images and dependencies to avoid future compatibility issues. Furthermore, for developers wishing to increase adoption of their technologies, we agree with Lin et al.~\cite{lin2020large} (2020) that the upward trend of lightweight application/language images suggest that ready-to-use images might encourage adoption.

For developers that write Dockerfiles to create images, we agree with the suggestions of Cito et al.~\cite{cito2017empirical}, Henkel et al.~\cite{henkel2020learning}, Lin et al.~\cite{lin2020large}, and Wu et al.~\cite{wu2020characterizing} that better tooling and education is needed to correct Dockerfiles that may produce non-optimal images. For instance, we find that there has been an increase in the Hadolint rule DL4006 where piping in a shell environment does not properly fail a Docker image build.

For the maintainers of Docker, they can observe the effects of deprecation efforts as well as the adoption of new features with our dataset, allowing further empirical reasoning about design decisions in the future which helps improve the overall software developer experience of using Docker. For example, we have observed that version pinning Dockerfile smells are actually increasing (compared to previous studies) which suggests that creating warnings about version pinning for a developer when building a Docker image would be beneficial.


We have included a replication package~\cite{msr2021} to further encourage future investigations on Dockerfile theories. Future works could extend the initial exploration of the Dockerfile characteristics presented in Section~\ref{df_state} to be a more in-depth analysis. Furthermore, additional definitions of \qq{signal and noise} when analyzing repositories could be considered. Although we mitigate the noise of analyzing project clones (forks) by Dockerfiles on a \qq{central repository} basis, other possible forms of signal and noise (e.g.\ engineered software projects vs homework assignments as defined by Munaiah et al.~\cite{munaiah2017curating}) have not been considered. Therefore, using our data in a future study, more emphasis can be placed on determining the definition of signal and noise when analyzing Dockerfiles to further confirm or disprove previous findings.

\bibliographystyle{IEEEtran}
\bibliography{main}


\end{document}